\documentclass[a4,11pt]{article}

\usepackage{enumerate}
\usepackage{graphicx}
\usepackage{times}
\usepackage{epsfig,longtable}
\usepackage{multicol}
\usepackage{color}
\usepackage{amssymb}
\usepackage{tabularx,array}
\usepackage{dcolumn}
\usepackage{eurosym}
\usepackage[T1]{fontenc}

\definecolor{linkcolor}{rgb}{0,0,0.6} 
\usepackage[pdftex,colorlinks=true, pdfstartview=FitV, linkcolor= linkcolor, citecolor= linkcolor, urlcolor= linkcolor, hyperindex=true,hyperfigures=true]{hyperref} 

\newcommand{\DUt}{\Delta U_\tau}
\newcommand{\Wt}{W_\tau}
\newcommand{\wt}{w_\tau}
\newcommand{\Qt}{Q_\tau}
\newcommand{\qt}{q_\tau}
\newcommand{\Wn}{W_n}
\newcommand{\wn}{w_n}
\newcommand{\Qn}{Q_n}
\newcommand{\qn}{q_n}
\newcommand{\om}{\omega}
\newcommand{\dd}{\mathrm{d}}
\newcommand{\Ieff}{I_{\mathrm{eff}}}
\newcommand{\ta}{\tau_\alpha}

\begin{document}
\title{\bf  Fluctuations in out of  equilibrium systems: \\
\ \ from theory to experiment }
\author{ S. Ciliberto,  S. Joubaud, A. Petrosyan \\ Universit\'e de Lyon \\
Ecole Normale Sup\'erieure de Lyon, Laboratoire de Physique ,\\
C.N.R.S. UMR5672,  \\ 46, All\'ee d'Italie, 69364 Lyon Cedex
07,  France\\}
\maketitle

\begin{abstract}

{
We introduce  from an experimental point
of view the main concepts of fluctuation theorems for work, heat
and entropy production in out of equilibrium systems. We will
discuss the important difference between the applications of these
concepts to stochastic systems and to a second class of systems
(chaotic systems) where the fluctuations are induced either by
chaotic flows or by fluctuating driving forces. We will mainly
analyze the stochastic systems using the measurements performed in
two experiments : a) a harmonic oscillator driven out of
equilibrium by an external force b) a colloidal particle trapped
in a time dependent  double well potential. We will rapidly
describe some consequences of fluctuation theorems and some useful
applications to the analysis of experimental data. As an example
the case of a molecular motor will be analyzed in some details.
Finally we will discuss the problems related to the
applications of fluctuation theorems to chaotic systems. }
\end{abstract}

\pagestyle{plain}

\noindent\hrulefill

\tableofcontents

\noindent\hrulefill

\section{Introduction}

\begin{figure}[h!]
\begin{center}
\includegraphics[width=1\linewidth]{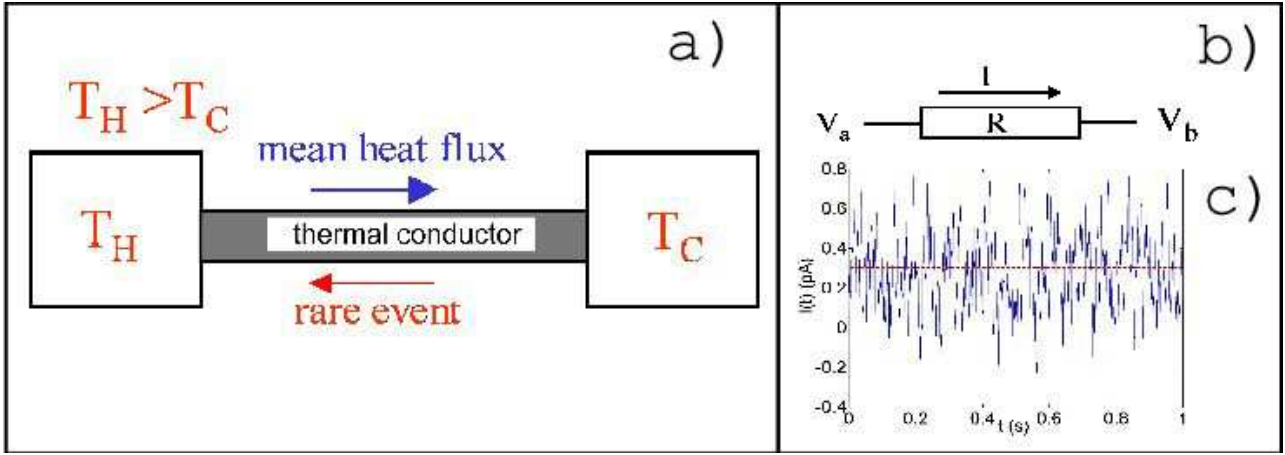}
\end{center}
\caption{ a) Schematic representation of a conductor whose
extremities are in contact with two heat baths at temperature $T_H$
and $T_C$ with $T_H> T_C$.  b) Electrical analogy. A conductor of
electrical resistance $R$ and kept at a temperature $T$ is
submitted to a potential difference $V=V_a-V_b$. c) Instantaneous
current $I$ flowing into the resistance using $R=10$~M$\Omega$,
$T=300$~K and $\tau_0=2$~ms.
  \label{fig:conduction} \small
}
\end{figure}

This article is a review of the main experimental applications of
Fluctuation Theorems (FTs) and summarizes the plenary talk given at
STATPHYS24. In order to define the main contents let us consider
several simple examples. The simplest and most basic out of
equilibrium system is a thermal conductor whose extremities are
connected to two heat baths at different temperatures, as sketched
in fig.\ref{fig:conduction}. The second law of thermodynamics
imposes that in average the heat flows from the hot to the cold
reservoir (from H to C in fig.\ref{fig:conduction}). However the
second law does not say anything about fluctuations and in
principle one can observe for a short time a heat current in the
opposite direction. What is the probability of observing these
rare events ? As a general rule when the size of the  system
decreases the role of fluctuations increases. Thus  from an
experimental point of view it is reasonable to think that such
rare events can be observed in systems that are small. A good
candidate could be  for example the thermal conduction in a
nanotube whose extremities are connected to two heat
baths \cite{prl:nanotube}, exactly in the spirit of
fig.\ref{fig:conduction}. In reality in this kind of  experiments
the measure of the mean quantities \cite{prl:nanotube} is already
difficult and of course the  analysis of fluctuations is even more
complicated. However there is an electrical analogy, shown in
fig.\ref{fig:conduction}b), of the thermal model of
fig.\ref{fig:conduction}a). Let us consider an electrical
conductor connected to a potential difference $V=V_A- V_B$ and
kept at temperature $T$ by a heat bath . If the mean current $\bar
I=V/R$ ($R$ being the electrical resistance of the conductor) is
of the order of $10^{-13}$~A and the injected power
is about $100 k_B T\simeq 10^{-19}$~J ($k_B$ is the
Boltzmann constant) then the instantaneous current inside the
resistance has fluctuations whose amplitude is  comparable to the
mean, as shown in fig.\ref{fig:conduction}c). The variance of
these fluctuations is $\delta I^2\simeq \ k_B T / (R \ \tau_0)$
where $\tau_0$ is the characteristic time constant of the
electrical circuit. In the specific case of Fig.
\ref{fig:conduction}c) the current  reverses with respect to the
mean value. The probability of having those negative currents have
been studied both theoretically and experimentally in
ref.~\cite{Cohen_electrical,Garnier} within the context of
fluctuation theorems, that we will present  in
sec.\ref{section_FT}.

{  We discuss a second example where the source of
fluctuation is not the coupling with a thermal bath, as in the
case of the electrical conductor,  but it is either a chaotic
flows or a chaotic force produced by the non-linear interaction of
many degrees of freedom of a dissipative system sustained by an
external driving.  We will refer to them as chaotic systems. Let
us consider a turbulent wind flowing around an object as sketched
in fig.\ref{fig:turbulence}a),b). The wind exerts a mean force
$F_0$ on the object but the instantaneous force, plotted in
fig.\ref{fig:turbulence}c),  is a strongly fluctuating quantity
which presents negative values \cite{Ciliberto_turbulence}, i.e.
the object moves against the wind,fig.\ref{fig:turbulence}b). In
such a case the mean work done by the wind on the object is about
$0.$~J$ \simeq 10^{20} k_B T$ and obviously thermal fluctuations do
not play any role but so does the chaotic flow, which produces the
fluctuations. Other similar examples can be found for example in
shaken granular media \cite{Menon,Joubaud_granular}, discussed in
sec.\ref{section:dynamical_system}.

\begin{figure}[h!]
\begin{center}
\includegraphics[width=1\linewidth]{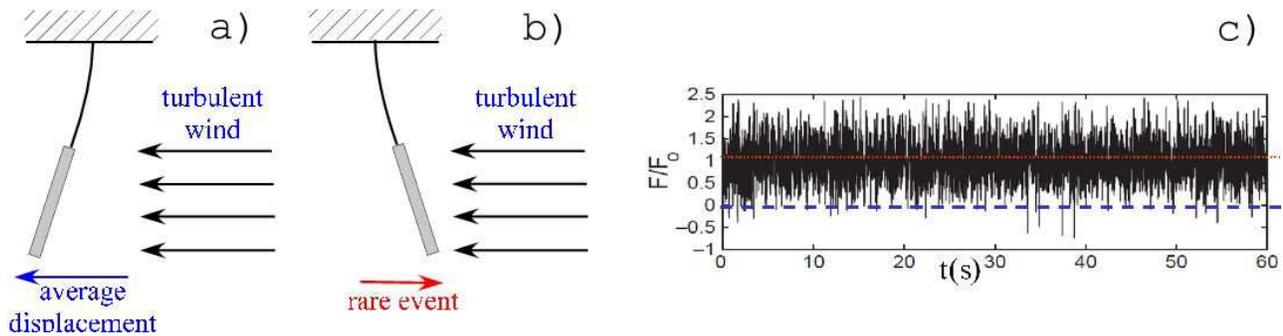}
\end{center}
\caption{ a)and b) Schematic representation of an object suspended
by an elastic beam and submitted to the pressure of a turbulent
wind a) average behavior b) rare event.  c) Time evolution of the
measured instantaneous force exerted by the turbulent wind on the
object. The details of this experiment can be found
\cite{Ciliberto_turbulence}
  \label{fig:turbulence} \small
}
\end{figure}

These examples  stress that in the two experiments the electrical
conductor and the turbulent wind we may observe the
counterintuitive effect that the instantaneous response of the
system is opposite to the mean value, in other words the system
has an instantaneous negative entropy production rate. This effect
is induced by the thermal fluctuations in the first case and by
the chaotic flow  in the second case. The question that we want
analyze in this article is whether the Fluctuations Theorem
(defined in section 3) is able to predict the probability of these
rare events in both cases, i.e. for the stochastic and the chaotic
systems. We will take an experimentalist approach and we will use
experimental results in order to introduce the main concepts.

The largest part of the article concerns stochastic systems
described by a Langevin dynamics. For  chaotic systems we will
mainly discuss the difficulty of comparing the  experimental
results with the theoretical predictions.} The article is
organized as follows. In section 2 we present the experimental
results on the energy fluctuations  measured in  a harmonic
oscillator driven out of equilibrium by an external force.
In section 3 the experimental results on the harmonic oscillator are used to
introduce the property of Fluctuation Theorems (FTs).
In section 4 the non linear case of a Brownian
particle confined in a time dependent double well potential is
presented. In section 5 we introduce the applications of the FT,
and as a more specific example we describe the measure of the
torque of a molecular motor. Finally in section 6 we discuss the
chaotic systems and we conclude in section 7.

\section{Work  and heat fluctuations in the harmonic oscillator}

The choice of discussing the dynamics of the harmonic oscillator
is dictated by the fact that it is relevant for many practical
applications such as the measure of the elasticity of
nanotubes\cite{Bellon_nano}, the dynamics of the tip of an AFM
\cite{Solano_AFM}, the MEMS and the thermal rheometer that we
developed several years ago to study the rheology of complex
fluids \cite{Bellon_Rheom_1,Bellon_Rheom_2}.

\subsection{The experimental set-up}

\begin{figure}
\centerline{\includegraphics[width=1\linewidth]{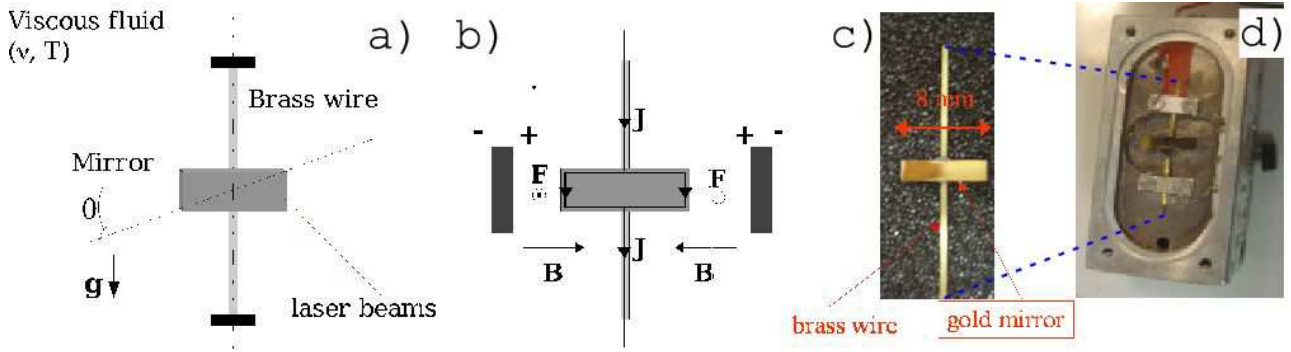}}
\caption{a) The torsion pendulum. b) The magnetostatic forcing.
c) Picture of the pendulum. d) Cell where the pendulum is installed.
} \label{fig:pendulum}
\end{figure}

This device is a very sensitive torsion pendulum as sketched in
fig.\ref{fig:pendulum}a). It is composed by a brass wire (length
$10$~$\textrm{mm}$, width $0.5$~$\textrm{mm}$, thickness $50$~$\mu
\textrm{m}$) and a glass mirror with a golden surface,
fig.\ref{fig:pendulum}c). The mirror  (length $2$~$\textrm{mm}$,
width $8$~$\textrm{mm}$, thickness $1$~$\textrm{mm}$) is  glued in
the middle of the brass wire. The elastic torsional stiffness of
the wire is $C = 4.65 \cdot 10^{-4}$ $\textrm{N.m.rad}^{-1}$. It
is enclosed in a cell, fig.\ref{fig:pendulum}d), which is filled
by a fluid. We used either air or a water-glycerol mixture at $60
\%$ concentration. The system is a harmonic oscillator with
resonant frequency $f_o=\sqrt{C/\Ieff}/(2\pi)=\omega_0/(2\pi)$ and
a relaxation time $\ta=2\Ieff/\nu=1/\alpha$. $\Ieff$ is the total
moment of inertia of the displaced masses ({\em i.e.} the mirror
and the mass of displaced fluid) \cite{Lamb}. The damping has two
contributions : the viscous damping $\nu$ of the surrounding fluid
and the viscoelasticity of the brass wire.

The angular displacement of the pendulum $\theta$ is measured by a
differential
interferometer~\cite{interferometer,Douarche05,Douarche06,Douarche04} which uses
the two laser beams reflected by the mirror
fig.\ref{fig:pendulum}a).  The measurement noise is two orders of
magnitude smaller than thermal fluctuations of the pendulum.
$\theta(t)$ is acquired with a resolution of $24$ bits at a
sampling rate of $8192$~$\textrm{Hz}$, which is about 40 times
$f_o$. We drive the system out-of-equilibrium by forcing it with
an external torque $M$ by means of a small electric current $J$
flowing in a coil glued behind the mirror
(Fig.~\ref{fig:pendulum}b). The coil is inside a static magnetic
field. The displacements of the coil and therefore the angular
displacements of the mirror are much smaller than the spatial
scale of inhomogeneity of the magnetic field. So the torque is
proportional to the injected current : $M = A.J$ ; the slope $A$
depends on the geometry of the system. The practical realization
of the montage is shown in
figs.~\ref{fig:pendulum}c),~\ref{fig:pendulum}d). In equilibrium
the variance $\delta \theta ^2$ of the thermal fluctuations of
$\theta$ can be obtained from equipartition, i.e. $\delta \theta =
\sqrt{ k_B \ T / C}\simeq 2$~nrad for our pendulum, where $T$ is
the temperature of the surrounding fluid.

\subsection{The equation of motion}

The dynamics  of the torsion pendulum can be assimilated to that
of a harmonic oscillator damped by the viscoelasticity of the
torsion wire and the viscosity of the surrounding fluid, whose
equation of motion reads in the temporal domain
\begin{equation}
   I_{\mathrm{eff}}\,\ddot{\theta} +
   \int_{-\infty}^{t} G(t-t')\, \dot{\theta}(t') \dd t' + C \theta = M+\eta,
   \label{eqofmotion}
\end{equation}
where $G$ is the memory kernel and $\eta$  the thermal noise. In
Fourier space (in the frequency range of our interest) this
equation takes the simple form
\begin{eqnarray}
   [- I_{\mathrm{eff}}\,{\om}^2 + \hat{C}]\, \hat{\theta} = \hat{M}+^\eta ,
\end{eqnarray}
where $\hat{\cdot}$ denotes the Fourier transform and $\hat{C} = C
+ i [C_1'' + \om C_2'']$ is the complex frequency-dependent
elastic stiffness of the system. $C_1''$ and $C_2''$ are  the
viscoelastic and viscous components of the damping term.

\begin{figure}[!h]
{\centerline{\includegraphics[width=1\linewidth]{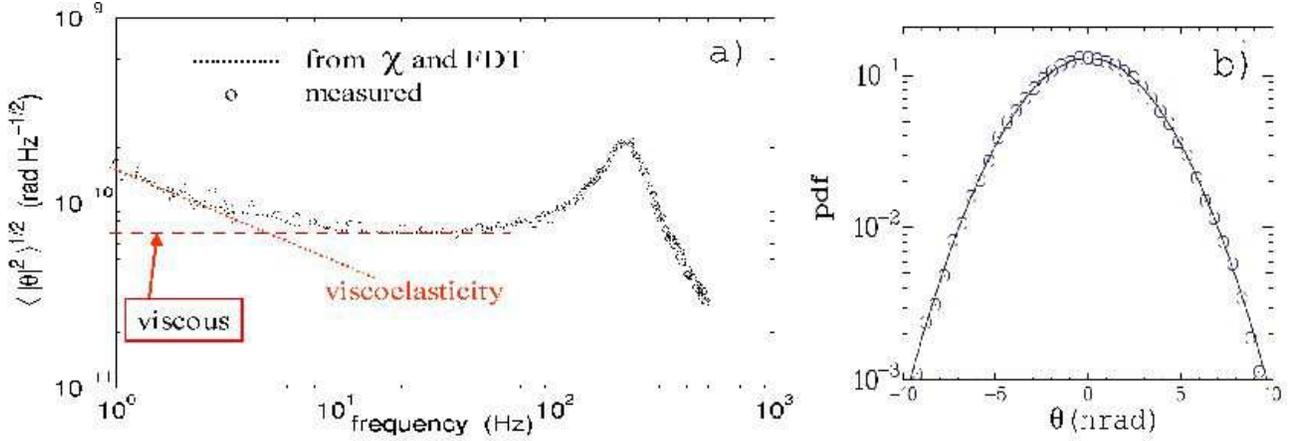}}}
\caption{Equilibrium: The pendulum  inside a glycerol-water
mixture with $M=0$. a) Square root of the power spectral density
of $\theta$. $\circ $ directly measured spectrum, black dotted
line is the spectrum estimated from the measure of $\chi$ and
using eq.\ref{fdt} The red dashed and dotted lines show the
viscous and viscoelastic component of the damping respectively. b)
Probability density function of $\theta $.  The continuous line is
a Gaussian fit} \label{fig:spectrum}
\end{figure}

\subsubsection{\it Equilibrium}
At equilibrium,{\em i.e.} $M=0$, the Fluctuation Dissipation Theorem
(FDT) gives a relation between the amplitude of the thermal
angular fluctuations of the oscillator and its response function.
The response function of the system $\hat{\chi} = \hat{\theta} /
\hat{M}= \frac{\hat \theta}{A \hat J}$ can be measured by applying
a torque with a white spectrum. When $M = 0$, the amplitude of the
thermal vibrations of the oscillator is related to its response
function via the fluctuation-dissipation theorem (FDT). Therefore,
the thermal fluctuation power spectral density (psd) of the
torsion pendulum reads for positive frequencies
\begin{equation}
   \langle { \vert \hat{\theta} \vert }^2 \rangle
   = \frac{4 k_B T}{\om} \, \mathrm{Im} \, \hat{\chi}
   =\frac{4 k_B T}{\om} \frac{C_1'' + \om \, C_2''}
   {{\lbrack -I_{\mathrm{eff}}\,{\om}^2 + C \rbrack}^2 + [C_1'' + \om \, C_2'']^2}.
   \label{fdt}
\end{equation}
The brackets are ensemble averages.
As an example, the spectrum of
$\theta$ measured in the glycerol-water mixture is shown in fig.\ref{fig:spectrum}a).
In this case the resonance frequency is $f_o=\sqrt{C/\Ieff}/(2\pi)=\omega_0/(2\pi)=217$~$\textrm{Hz}$ and
the  relaxation time $\ta=2\Ieff/\nu=1/\alpha= 9.5$~$\textrm{ms}$
The measured
spectrum is compared with that obtained from eq.\ref{fdt} using
the measured $\chi$. The viscoelastic component at low frequencies
correspond to a  constant $C_1''\ne 0$. Indeed if $\om \rightarrow
0$ then from eq.\ref{fdt} $\langle { \vert \hat{\theta} \vert }^2
\rangle \propto 1/\om $ as seen in fig.\ref{fig:spectrum}a).
Instead if $C_1''=0$ then for $\om \rightarrow 0$ from
eq.\ref{fdt} the spectrum is constant as a function of $\omega$.
It is important to stress that in the viscoelastic case the noise
$\eta$ is correlated and the process is not Markovian, whereas in
the viscous case the process is Markovian. Thus by changing the
quality of the fluid surrounding the pendulum one can tune the
Markovian nature of the process. In the following we will consider
only the experiment in the glycerol-water mixture where the
viscoelastic contribution is visible only  at very low frequencies
and is therefore negligible. This allows a more precise comparison
with theoretical predictions often obtained for Markovian
processes. The probability density function (pdf) of $\theta$,
plotted in fig.\ref{fig:spectrum}b), is a Gaussian.

\subsection{Non-equilibrium Steady State (NESS): Sinusoidal forcing\label{sec:sinusexp}}
\begin{figure}[!h]
\centerline{\includegraphics[width=1.0\linewidth]{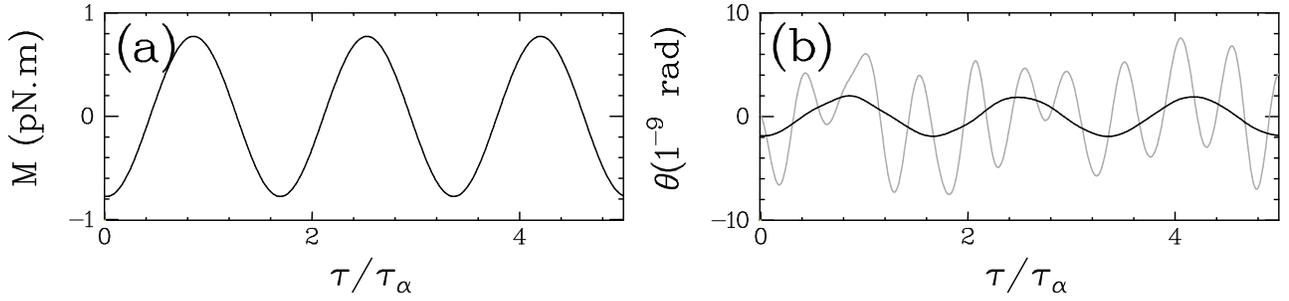}}
\caption{a) Sinusoidal driving torque applied to the oscillator.
b) Response of the oscillator to this periodic forcing (gray line)
; the dark line represents the mean response $\langle \theta(t)
\rangle$.} \label{fig:sinustorque}
\end{figure}

We now consider a periodic forcing of amplitude $M_o$ and frequency $\omega_d$, i.e. $M(t) = M_o \sin(\omega_d t)$ \cite{Douarche06}-\cite{Joubaud2007}.  This
is a very common kind of forcing which has been already studied in
the case of the first order Langevin equation
\cite{Blickle} and of the two level system \cite{Schuler}
and in a different context for the second order Langevin equation
\cite{Zamponi}. Furthermore this is a very general case because
using Fourier transform, any periodical forcing can be decomposed
in a sum of sinusoidal forcing. We explain here the behavior of a
single mode. Experiments have been performed at various $M_o$ and
$\omega_d$. We present here the results for a particular amplitude
and frequency: $M_o = 0.78$~$\textrm{pN.m}$ and $\omega_d/(2\pi) =
64$~$\textrm{Hz}$. This torque is plotted in
Fig.~\ref{fig:sinustorque}a. The mean of the response to this
torque is sinusoidal, with the same frequency, as can be seen in
Fig.~\ref{fig:sinustorque}b. The system is clearly in a
non-equilibrium steady state (NESS).

The work done by the torque $M(t)$ on a time $\tau_n=2\pi \ n /\om_d$
is
\begin{equation}
W_n = W_{\tau=\tau_n} =  \ \int_{t_i}^{t_i+\tau_n} M(t) {d \theta
\over dt} dt
\end{equation}

\begin{figure}[!h]
{\includegraphics[width=1.0\linewidth]{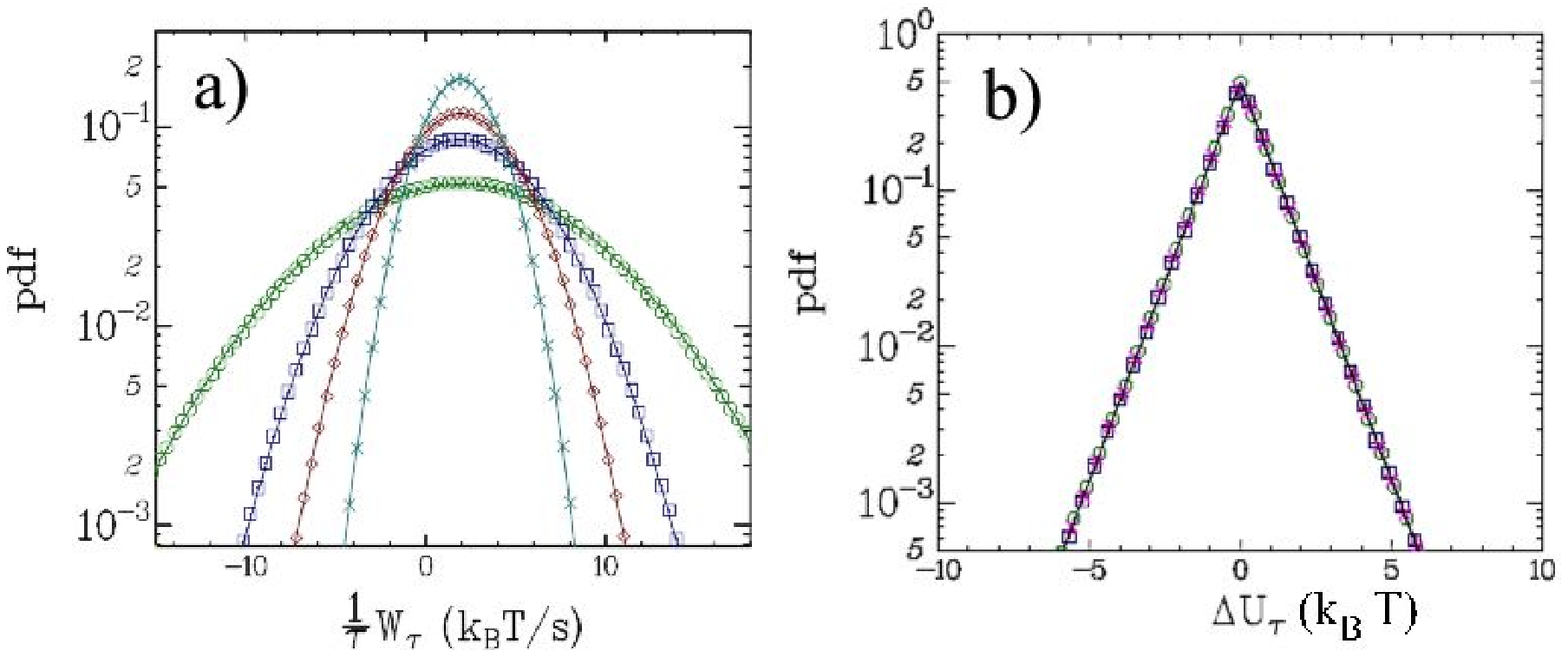}}
\caption{Sinusoidal forcing. a) Pdf of $\Wt$;  $n=7$ ($\circ$),
$n=15$ ($\Box$), $n=25$ ($\diamond$) and $n=50$ ($\times$). b) Pdf of
$\DUt$} \label{fig:pdf_w_U}
\end{figure}

As $\theta$ fluctuates also $W_n$ is a fluctuating quantity whose
probability density function (pdf) is plotted in
fig.~\ref{fig:pdf_w_U}a) for various $n$.  This plot has
interesting features. Specifically, work fluctuations are Gaussian
for all values of $n$ and  $\Wt$ takes negative values as long as
$\tau_n$ is not too large.
The probability of having negative values of $\Wt$ decreases when $\tau_n$ is increased.
There is a finite probability of having
negative values of the work, in other words the system may have an
instantaneous negative entropy production rate although the
average of the work $<W_n>$ is of course positive ($<.>$ stands
for ensemble average).  In this specific example is $<W_n>=0.04 \
n (k_B \ T)$. We now consider the energy balance for the system.

\subsection{Energy balance}

As the fluid is rather viscous we will take into account only the
standard viscosity that is $C_1''=0$ and $C_2''=\nu$. In such a
case eq.\ref{eqofmotion} simplifies
\begin{equation}
\Ieff\,\frac{\dd^2{\theta}}{\dd t^2}+ \nu \,\frac{\dd{\theta}}{\dd
t}  + C\,\theta = M +  \ \eta, \label{eq:Langevin_oscillator}
\end{equation}
where $\eta$ is the thermal noise, which in this case is
delta-correlated in time: $<\eta(t) \ \eta(t')>={2\ k_B \ T\ \nu}
\delta(t-t')$.

When the system is driven out of equilibrium using a deterministic
torque, it receives some work and a fraction of this energy is
dissipated into the heat bath. Multiplying
Eq.~(\ref{eq:Langevin_oscillator}) by $\dot{\theta}$ and
integrating between $t_i$ and $t_i+\tau$, one obtains a
formulation of the first law of thermodynamics between the two
states at time $t_i$ and $t_i+\tau$
(Eq.~(\ref{eq:Energy_conservation})). This formulation has been
first proposed in ref.\cite{Sekimoto} and used in other
theoretical and experimental works \cite{Sasa,Blickle}. The change
in internal energy $\DUt$ of the oscillator over a time $\tau$,
starting at a time $t_i$, is written as:
\begin{equation}
\DUt = U(t_i+\tau) - U(t_i) =  \Wt-\Qt
\label{eq:Energy_conservation}
\end{equation}
where $\Wt$ is the work done on the system over a time $\tau$ :
\begin{equation}
\Wt =  \int_{t_i}^{t_i+\tau} M(t') \frac{\dd\theta}{\dd t}(t') \dd
t' \label{eq:Wdef}
\end{equation}
and $\Qt$ is the heat dissipated by the system.   The internal
energy is the sum of the potential energy and the kinetic energy :
\begin{equation}
U(t)= \left\{\frac{1}{2} \Ieff\left[ \frac{\dd \theta}{\dd t}(t)
\right]^{2} +\frac{1}{2} C \theta(t)^2 \right\}. \label{eq:Udef}
\end{equation}
The heat transfer $\Qt$ is deduced from equation
(\ref{eq:Energy_conservation}) ; it has two contributions :
\begin{eqnarray}
\Qt &=& \Wt -\DUt \nonumber\\
&=&  \int_{t_i}^{t_i+\tau} \nu \left[ \frac{\dd \theta}{\dd t}
(t')\right]^{2}\dd t' -  \int_{t_i}^{t_i+\tau} \eta(t') \frac{\dd
\theta}{\dd t}(t') \dd t' . \label{eq:Qdef}
\end{eqnarray}
The first term corresponds to the viscous dissipation and is
always positive, whereas the second term can be interpreted as the
work of the thermal noise which has a fluctuating sign. The
second law of thermodynamics imposes $\langle \Qt \rangle$ to be
positive.

\begin{figure}[!h]
{\includegraphics[width=1.0\linewidth]{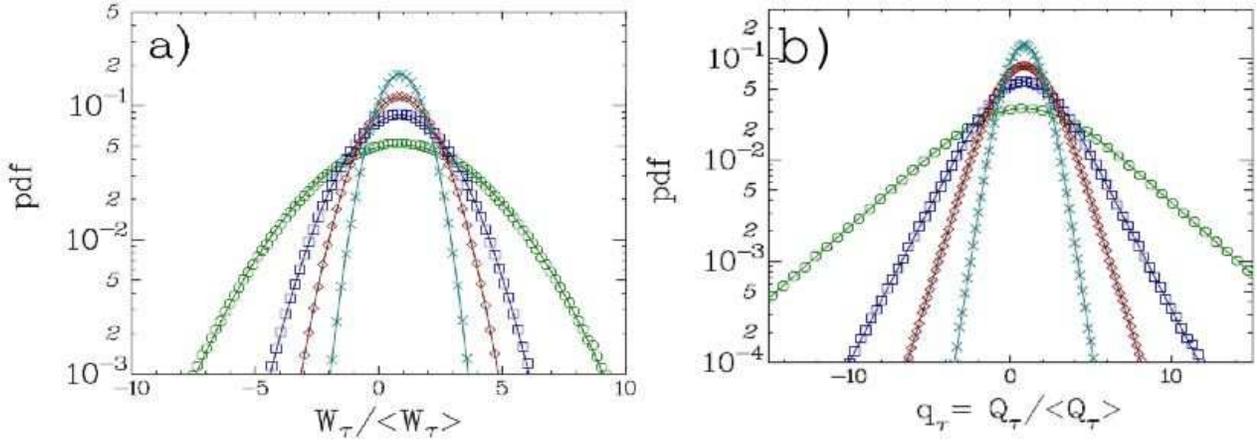}}
\caption{Sinusoidal forcing. a) Pdf of $W_\tau$  b) Pdf of
$Q_\tau$ for various $n$:   $n=7$ ($\circ$), $n=15$ ($\Box$), $n=25$
($\diamond$) and $n=50$ ($\times$). The continuous lines in this
figures are not fits but are analytical predictions obtained from
the Lnagevin dynamics as discussed in
sect.\ref{sec:theoretical_results}} \label{fig:compare_pdf_w_q}
\end{figure}

\subsection{Heat fluctuations}
The dissipated heat $\Qt$ can not be directly measured  because we
have seen that eq.\ref{eq:Qdef} contains  the work of the noise
(the heat bath) that experimentally is impossible to measure,
because $\eta$ is unknown. However $\Qt$ can be obtained
indirectly from the measure of $\Wt$ and $\DUt$, whose pdf
measured during the periodic forcing  are  exponential for any
$\tau$, as shown in fig.\ref{fig:pdf_w_U}b. We first do some
comments on the average values. The average of $\DUt$ is obviously
vanishing because the time $\tau$ is a multiple of the period of
the forcing. Therefore $\langle \Wn \rangle$ and $\langle \Qn
\rangle$ are equal.

We rescale the work $\Wt$ (the heat $\Qt$) by the average work
$\langle \Wt \rangle$ (the average heat $\langle \Qt \rangle$) and
define: $\wt = \frac{\Wt}{\langle \Wt \rangle}$ ($\qt =
\frac{\Qt}{\langle \Qt \rangle}$). In the present article, $x_\tau$, respectively $X_\tau$, stands
for either $\wt$ or $\qt$, respectively $\Wt$ or $\Qt$.

We compare now the pdf of $\wt$ and $\qt$ in
Fig.~\ref{fig:compare_pdf_w_q}. The pdfs of heat fluctuations
$\qn$ have exponential tails (Fig.~\ref{fig:compare_pdf_w_q}b). It
is interesting to stress that although the two variables $\Wt$ and
$\Qt$ have the same mean values they have a very different pdf.
The pdf of $\wt$ are gaussian  whereas those of $\qt$ are
exponential. On a first approximation the pdf of $\qt$ are the
convolution of a Gaussian (the pdf of $\Wt$) and exponential (the
pdf of $\DUt$). In Figs.~\ref{fig:compare_pdf_w_q} the continuous
lines are analytical predictions obtained from the Langevin
dynamics with no adjustable parameter (see
sect.\ref{sec:theoretical_results}).


\section{Fluctuation theorem \label{section_FT}}
In the previous section we have seen that both $\Wt$ and $\Qt$
present negative values, {\em i.e.}  the second law is
verified only on average but the entropy
production can have instantaneously negative values.
The
probabilities of getting positive and negative entropy production
are quantitatively related in non-equilibrium systems by the
Fluctuation Theorem (FTs).

There are two classes of FTs. The {\it Stationary State
Fluctuation Theorem} (SSFT) considers a non-equilibrium steady
state. The {\it Transient Fluctuation Theorem}
(TFT) describes transient non-equilibrium states where $\tau$
measures the time since the system left the equilibrium state.
A Fluctuation Relation (FR) examines the symmetry around $0$ of
the probability density function (pdf) $p(x_\tau)$ of a quantity
$x_\tau$, as defined in the previous section.
It compares the probability to have a positive event
($x_\tau = +x$) versus the probability to have a negative event
($x_\tau = -x$). We quantify the FT using a function $S$ (symmetry
function) :
\begin{equation}
S(x_\tau) = \frac{k_B \ T} {\langle X_\tau \rangle}\ln \left(
\frac{p(x_\tau=+x)}{p(x_\tau=-x)}\right). \label{eq:FT}
\end{equation}

The {\it Transient Fluctuation Theorem} (TFT) states that the
symmetry function is linear with $x_\tau$ for any values of the
time integration $\tau$ and  the proportionality coefficient is
equal to $1$ for any value of $\tau$.
\begin{equation}
S(x_\tau)=x_\tau, \quad \forall x_\tau, \quad \forall \tau.
\label{eq:TFT}
\end{equation}
Contrary to TFT, the {\it Stationary State Fluctuation Theorem}
(SSFT) holds only in the limit of infinite time ($\tau$).
\begin{equation}
\lim_{\tau \rightarrow \infty} S(x_\tau) = x_\tau.  \label{eq:SSFT}
\end{equation}

In the following we will assume linearity at finite time $\tau$
\cite{Cohen_electrical,Cohen} and use the following general
expression :
\begin{equation}
S(x_\tau)= \ \Sigma_x(\tau) \ x_\tau
\label{eq:FTs}
\end{equation}
where for SSFT  $\Sigma_x(\tau)$ takes into account the finite time corrections
and $\lim_{\tau \rightarrow \infty} \Sigma_x(\tau)=1$ whereas
$\Sigma_x(\tau)=1, \quad \forall \ \tau $ for TFT.

However these claims are not universal because they depend on the kind of $x_\tau$ which is used.
Specifically we will see in the next sections  that the results are  not exactly  the same if $X_\tau$ is replaced by any one of  $\Wt$, $\Qt$ and $(T \ s_{tot,\tau})$, defined in sect.\ref{section_FT_harmonic}. Furthermore the definitions given in this section are appropriate for
stochastic systems and in sect.\ref{section:dynamical_system} we will discuss
the differences between stochastic and chaotic systems.

\subsection{Short history of FTs}

The first numerical evidence of relations of this kind has been
given by Evans et  al. in ref.\cite{Evansetal93} whereas the TFT
was proved in ref.\cite{Evans-Searles}. In 1995 Gallavotti and
Cohen \cite{GallavottiCohen95} proved SSFT for dynamical systems
although in such a case $x_\tau$ takes a different meaning that we
will discuss in sect.\ref{section:dynamical_system}. The proof of
SSFT has been extended to  stochastic dynamics in ref.
\cite{Cohen_electrical,Lebowitz,Kurchan98,Farago,Cohen}.
Furthermore van Zon and Cohen proved that there is an important
difference between the FTs for the injected power and those for
the dissipated power~\cite{Cohen_electrical,Cohen}. The SSFT has
been  proved also for other quantities such as the dissipation
function \cite{Searles_2007} and the total entropy
\cite{Seifert2005,Puglisi2006}. Other theoretical papers studied
FT and  the reader may find a review in
ref.\cite{Rondoni2007,Zamponi_comment}. Experiments searching for
FTs have been performed in dynamical
systems~\cite{Ciliberto_turbulence,Menon,Ciliberto98}, but
interpretations are very difficult because a quantitative
comparison with theoretical prediction can be doubtful. Other
experiments have been performed in stochastic systems described by
a first order Langevin equation: a Brownian particle in a moving
optical trap~\cite{Wangetal:02:05} and an out-of-equilibrium
electrical circuit~\cite{Garnier} in which existing theoretical
predictions~\cite{Cohen_electrical,Cohen} were verified. Other
experimental tests for FTs have been performed on driven two level
systems \cite{Schuler} and on colloids \cite{Blickle}.

\subsection{ FTs for Gaussian variables \label{sec:gaussian_FT}}

Let us suppose the the variable $X_\tau$ has a Gaussian
distribution of mean $<X_\tau>$ and variance $\sigma_{X_\tau}^2$.
It is easy to show that in order to satisfy FTs, the variable
$X_\tau$ must have the following  statistical property:
\begin{equation}
\sigma_{X_\tau}^2 = 2 \ k_B \ T \ <X_\tau>
\label{eq:Gaussian_FT}
\end{equation}
This is an interesting relation because it imposes  that the
relative  fluctuations of $X_\tau$ are
\begin{equation}
{\sigma_{X_\tau} \over X_\tau } = \sqrt{ {2 \ k_B \ T \over  <X_\tau>} }
\label{eq:relative_fluct}
\end{equation}

This means that the probability of having negative events reduces by increasing  $X_\tau$, specifically from eq.\ref{eq:relative_fluct} it follows that $P(X_\tau<0)=\rm{erfc}(\sqrt{<X_\tau>/(2 \ k_B T)})$ where
$\rm{erfc}$ is the complementary error function.  It is now possible to estimate the length of the time interval $t_{obs}$ needed to observe  at least one negative event, which is:
\begin{equation}
t_{obs}={\tau\over \rm{erfc}\left(\sqrt{<X_\tau> \over 2 \ k_B \ T }\ \right)}
\end{equation}
where we used the fact that all the  values $\Wt$ computed on
different intervals of length $\tau$ are independent, which is
certainly true if $\tau$ is larger than the correlation time.

Let us consider the specific
example of section \ref{sec:sinusexp}, i.e. $<\Wt>=0.04 n (k_BT)$
at  $\omega_d/(2 \ \pi)= 67$~Hz, $M_o =
0.78$~$\textrm{pN.m}$   and  $\tau=2 \pi n / \omega_d $. The
pdf of $\Wt$ are Gaussian in this case (Fig.\ref{fig:pdf_w_U})
and, as we will see in the next section, they satisfy SSFT for
large $\tau$. Therefore eq.\ref{eq:relative_fluct} holds for
$X_\tau=\Wt$ and we may estimate $t_{obs}$ in the asymptotic limit $\ta <<\tau$.  For example at $n=200$, one obtains from the above mentioned experimental values $\tau \simeq 3s>>\ta$
and   $\ <\Wt> = 8 k_BT $. Inserting these experimental values  in  eq.\ref{eq:relative_fluct} one gets roughly a negative event over an observational
 time $t_{obs}\simeq 641 s$, which is already a rather long time for the distance between two
 events. For larger $n$ and larger $M_0$ this time becomes exponentially large.
 This justifies the fact that millions of data are necessary in order to
 have a reliable measure of SSFT.

\subsection{ FTs for $\Wt$ and $\Qt$ measured in
the harmonic oscillator \label{section_FT_harmonic}}

The questions we ask are whether for finite time FTs are satisfied
for either $x_\tau = w_\tau$ or $x_\tau = q_\tau$  and what are
the finite time corrections. In a first time, we test the
correction to the proportionality between the symmetry function
$S(x_\tau)$ and $x_\tau$. In the region where the symmetry
function is linear with $x_\tau$, we define the slope
$\Sigma_x(\tau)$, i.e. $S(x_\tau) = \Sigma_x(\tau) x_\tau$. In a
second time we measure finite time corrections to the value
$\Sigma_x(\tau) = 1$ which is the asymptotic value expected from
FTs.

In this review article we will focus on the SSFT applied to the experimental results
of sect. \ref{sec:sinusexp} and to other examples. The TFT will be not discussed here and the interested readers may look at ref.\cite{Joubaud2007}.

\begin{figure}[!h]
{\includegraphics[width=1.0\linewidth]{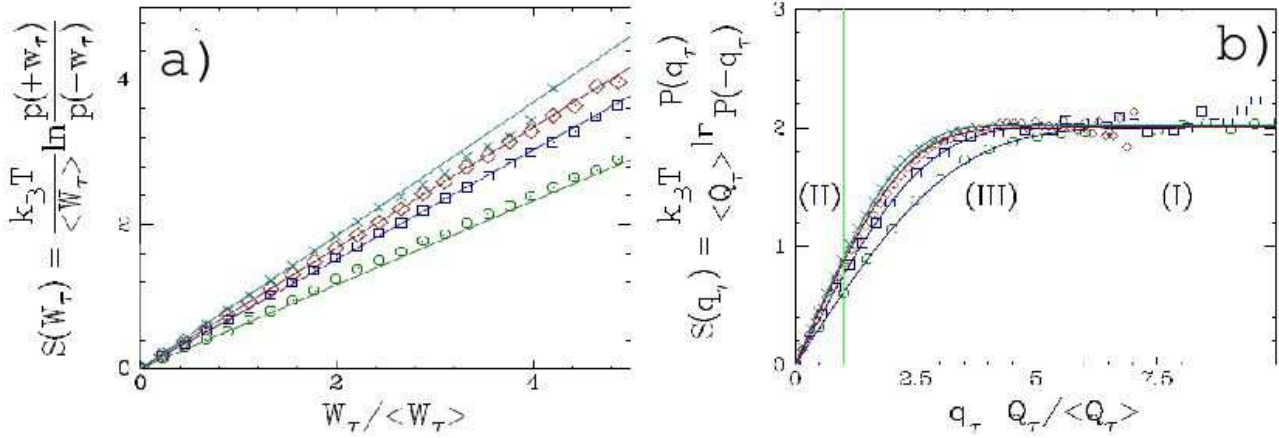}}
\caption{Sinusoidal forcing. Symmetry functions for SSFT.  a)
Symmetry functions $S(\wt)$  plotted as a function of $\wt$ for
various $n$:   $n=7$ ($\circ$), $n=15$ ($\Box$), $n=25$ ($\diamond$) and
$n=50$ ($\times$).  For all $n$ the dependence of $S(\wt)$  on
$\wt$ is linear, with slope $\Sigma_w(\tau)$. b) Symmetry
functions $S(\qt)$ plotted as a function of $\qt$ for various $n$.
The dependence of $S(\qt)$  on $\qt$ is linear only
for $\qt <1$. Continuous lines are is
theoretical predictions.} \label{fig:S_W_S_Q}
\end{figure}

\begin{figure}[!h]
{\centerline{\includegraphics[width=0.6\linewidth]{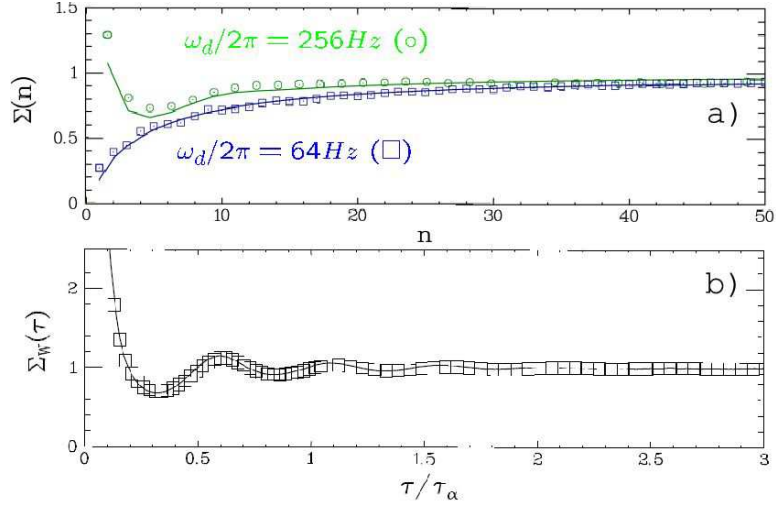}}}
\caption{ Finite  time corrections for SSFT. a) Sinusoidal forcing. $\Sigma_w(\tau)$ as a function of n obtained from the slopes of the straight lines of Fig.\ref{fig:S_W_S_Q}a) ($\square $). The circles correspond to another measurement performed at a different frequency. The finite time corrections depend on the driving frequency. The slope $\Sigma_q(\tau)$  measured for $\qt<1$  (Fig.\ref{fig:S_W_S_Q}b) have  exactly the same values of $S_w(\tau)$ as a function of $n$.  b) Linear forcing. $\Sigma_w(\tau)$ measured as a function of $\tau$ with the driving torque $M$ has a linear dependence on time. The finite time corrections depend on form of the driving. }
\label{fig:compare_Sigma}
\end{figure}

From the pdfs of $\wt$ and $\qt$ plotted in
Figs.\ref{fig:compare_pdf_w_q}, we compute the symmetry functions
defined in eq.\ref{eq:FT}. The symmetry  function  $S(\wn)$ are
plotted in Fig.~\ref{fig:S_W_S_Q}a) as a function of $\wn$.
They are  linear in $\wn$. The slope $\Sigma_w(n)$ is not equal
to $1$ for all $n$ but there is a correction at finite time
(Fig.~\ref{fig:compare_Sigma}a). Nevertheless, $\Sigma_w(n)$ tends
to $1$ for large $n$. Thus SSFT is satisfied for $\Wt$ and for a
sinusoidal forcing.  The convergence is very slow and we have to
wait a large number of periods of forcing for the slope to be $1$
(after $30$ periods, the slope is still $0.9$). This behavior is
independent of the amplitude of the forcing $M_o$ and consequently
of the mean value of the work $\langle W_n \rangle$, which, as
explained in sec.\ref{sec:gaussian_FT}, changes only the time
needed to observe a negative event. The system satisfies the SSFT
for all forcing frequencies $\omega_d$ but finite time corrections
depend on $\omega_d$, as can be seen in
Fig.~\ref{fig:compare_Sigma}a).

We now analyze the pdf of $\qt$ (Fig.\ref{fig:compare_pdf_w_q}b)) and we compute the
 symmetry functions $S(\qn)$ of $\qn$ plotted
in Fig.~\ref{fig:S_W_S_Q}b) for different values of $n$. They are
clearly very different from those of $\wn$ plotted in
Fig.~\ref{fig:S_W_S_Q}a).  For $S(\qn)$ three different regions
appear:

(I) For large fluctuations $\qn$, $S(\qn)$ equals $2$. When $\tau$
tends to infinity, this region spans from $\qn = 3$ to infinity.

(II) For small fluctuations $\qn$, $S(\qn)$ is a linear function
of $\qn$. We then define $\Sigma_q(n)$ as the slope of the
function $S(\qn)$, {\it i.e.} $S(\qn) = \Sigma_q(n) \, \qn$.
We have measured \cite{Joubaud2007} that $\Sigma_q(n)= \Sigma_w(n)$ for all the values of $n$, i.e. finite time corrections are the same for
heat and  work.  Thus $\Sigma_q(n)$
tends to $1$ when $\tau$ is increased and SSFT holds in this
region II which spans from $\qn=0$ up to $\qn = 1$ for large
$\tau$. This effect has been discussed for the first time in refs.\cite{Cohen,Cohen_electrical}.

(III) A smooth connection between the two behaviors.

These regions define the Fluctuation Relation from the heat
dissipated by the oscillator. The limit for large $\tau$ of the
symmetry function $S(\qt)$ is rather delicate and it has been discussed in ref.\cite{Joubaud2007}.

The conclusions of this experimental analysis is that SSFT holds
for work for any value of $\wt$ whereas for heat it holds only for
$\qt <1$. The finite time correction to FTs, described by
$1-\Sigma$ are not universal. They are the same both for $\wt$ and
$\qt$ but they depend on the driving frequency as shown in
Fig.\ref{fig:compare_Sigma}a). Furthermore they depend on the kind
of driving force. In Fig.~\ref{fig:compare_Sigma}b) we plot
$\Sigma_w(\tau)$ measured when the harmonic oscillator is driven
out of equilibrium by a linear ramp\footnote{The stationarity in
the case of  a ramp is discussed in
ref.\cite{Cohen,Joubaud2007}}. The difference with respect
Fig.~\ref{fig:compare_Sigma}a) is quite evident.

\subsection{Comparison with theory \label{sec:theoretical_results}}
This experimental analysis allows a very precise comparison with
theoretical predictions using the Langevin equation
(eq.\ref{eq:Langevin_oscillator}) and using two experimental
observations: a) the properties of heat bath are not modified by
the driving and b) the fluctuations of the $\Wt$ are Gaussian (see
also \cite{Speck}, where it is shown that in Langevin dynamics
$\Wt$ has a Gaussian distribution for any kind of deterministic
driving force if the properties of the bath are not modified by
the driving and the potential is harmonic). The observation in
point  a) is extremely important because it is always assumed to
be true in all the theoretical analysis. In ref.\cite{Joubaud2007}
this point has been precisely checked. Using these experimental
observations one can compute the pdf of $\qt$ and the finite time
corrections $\Sigma (\tau)$ to SSFT (see ref.\cite{Joubaud2007}).
The continuous lines in
Fig.~\ref{fig:compare_Sigma}, Fig.~\ref{fig:S_W_S_Q} and
Fig.~\ref{fig:pdf_w_U} are not fit but analytical predictions,
with no adjustable parameters, derived from the Langevin dynamics
of eq.\ref{eq:Langevin_oscillator} (see  ref.\cite{Joubaud2007} for more
details).

\subsection{The trajectory dependent entropy \label{sect:trajectory_entropy}}
In previous sections we have studied the energy $\Wt$ injected
into the system in the time $\tau$ and the energy dissipated
towards the heat bath $\Qt$. These two quantities and the internal
energy are related by the first law of thermodynamics
(eq.\ref{eq:Qdef}).
\begin{figure}[!h]
{\centerline{\includegraphics[width=1\linewidth]{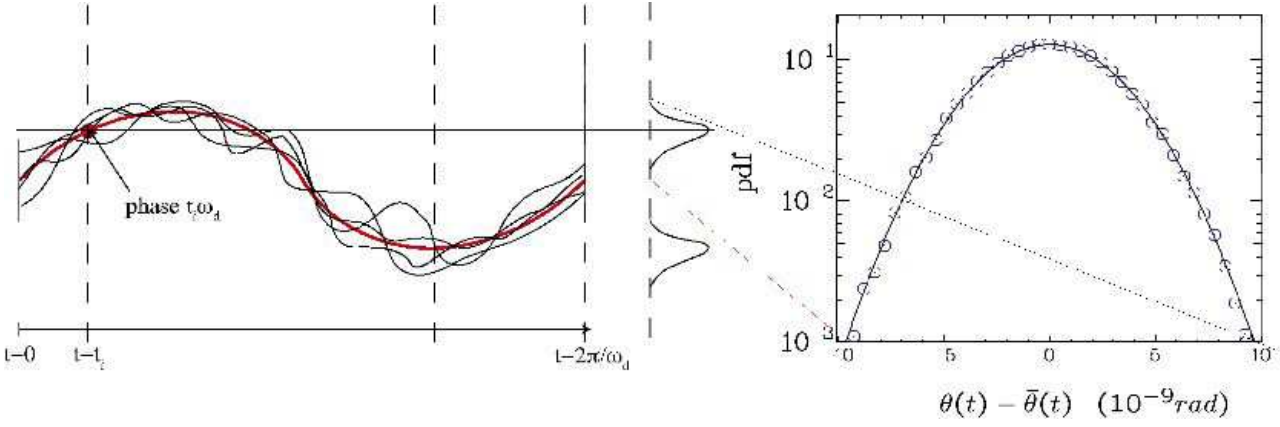}}}
\caption{a)  Schematic diagram illustrating the method to compute
the trajectory dependent entropy, b) Pdf of $\theta(t)$ around the mean trajectory $<\theta(t)>$. The continuous line
is the equilibrium distribution} \label{fig:trajectories}
\end{figure}
Following notations of ref~\cite{Seifert2005}, we define the
entropy variation in the system during a time $\tau$ as :
\begin{equation}
\Delta s_{\rm{m},\tau} = \frac{1}{T}Q_\tau.
\label{eq:def_medium_entropy}
\end{equation}
For thermostated
systems, entropy change in medium behaves like the dissipated
heat. The non-equilibrium Gibbs entropy is :
\begin{equation}
\langle s(t) \rangle=-k_B \int \dd \vec{x}
p(\vec{x}(t),t,\lambda_t) \ln p(\vec{x}(t),t,\lambda_t)
\end{equation}
where $\lambda_t$ denotes the set of control parameters at time
$t$ and $p(\vec{x}(t),t,\lambda_t)$ is the probability density
function to find the particle at a position $\vec{x}(t)$ at time
$t$, for the state corresponding to $\lambda_t$. This expression
allows the definition of a "trajectory-dependent" entropy :
\begin{equation}
s(t) \equiv -k_B \ln p(\vec{x}(t),t,\lambda_t)
\label{eq:def_Gibbs_entropy}
\end{equation}

\begin{figure*}[!ht]
\begin{center}
\includegraphics[width=0.8\linewidth]{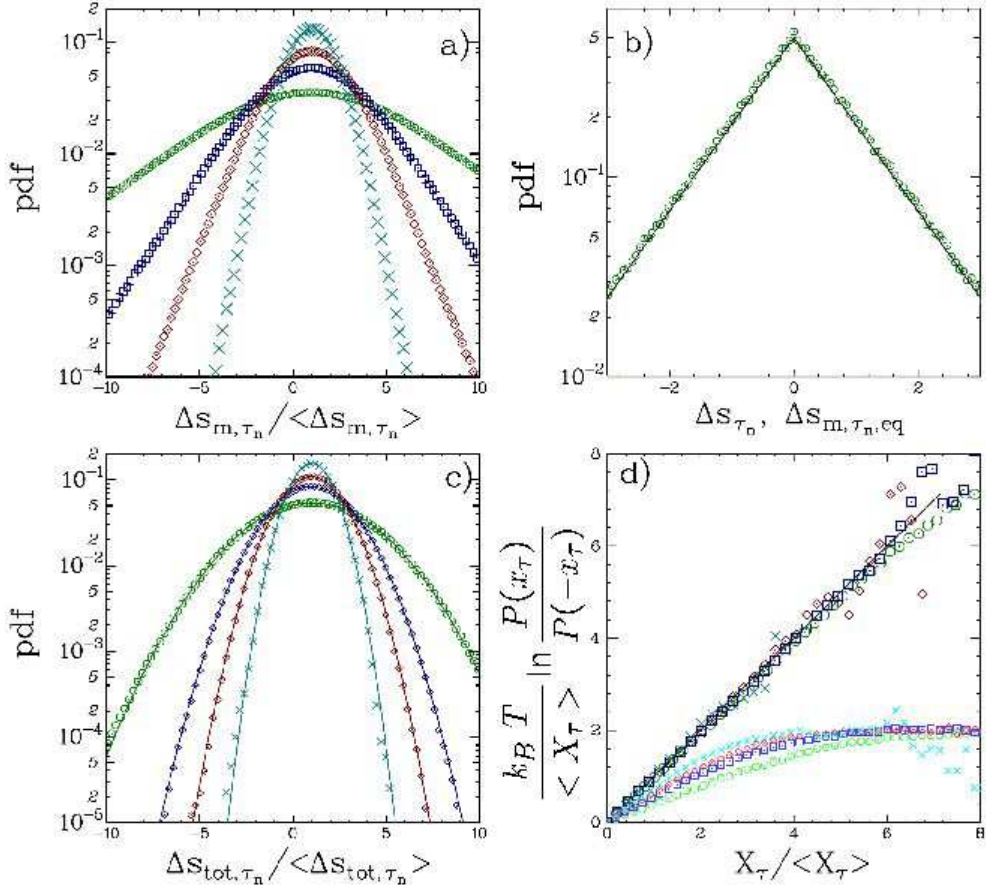}
\caption{Torsion pendulum. a) pdfs of the normalized entropy
variation $\Delta s_{\rm{m},\tau_n}/\langle \Delta
s_{\rm{m},\tau_n}\rangle$ integrated over $n$ periods of forcing,
with $n=7$ ($\circ$), $n=15$ ($\Box$), $n=25$ ($\diamond$) and
$n=50$ ($\times$). b) pdfs of $\Delta s_{\tau_n}$, the
distribution is independent of $n$ and here $n=7$. Continuous line
is the theoretical prediction for equilibrium entropy exchanged
with thermal bath $\Delta s_{\rm{m},\tau_n,eq}$. c) pdfs of the
normalized total entropy $\Delta s_{\rm{tot},\tau_n}/\langle
\Delta s_{\rm{tot},\tau_n}\rangle$, with $n=7$ ($\circ$), $n=15$
($\Box$), $n=25$ ($\diamond$) and $n=50$ ($\times$). d) Symmetry
functions for the normalized entropy variation in the system
(small symbols in light colors and $X_\tau$ stands for $T \ \Delta
s_{\rm{m},\tau_n}=\Qt$) and for the normalized total entropy
(large symbols in dark colors and $X_\tau$ stands for $T \ \Delta
s_{\rm{tot},\tau_n}$) for the same values of $n$.}
\label{pendule_sinus:pdf}
\end{center}
\end{figure*}
The variation $\Delta s_{\rm{tot},\tau}$ of the total entropy
$s_{\rm{tot}}$ during a time $\tau$ is the sum of the entropy
change in the system during $\tau$ and the variation of the
"trajectory-dependent" entropy in a time $\tau$, $\Delta
s_\tau\equiv s(t+\tau) - s(t)$ :
\begin{equation}
\Delta s_{\rm{tot},\tau} \equiv s_{\rm{tot}}(t+\tau) -
s_{\rm{tot}}(t) = \Delta s_{\rm{m},\tau} + \Delta s_\tau
\label{eq:def_total_entropy}
\end{equation}
In this section, we study fluctuations of $\Delta
s_{\rm{tot},\tau}$ computed using (\ref{eq:def_medium_entropy})
and (\ref{eq:def_Gibbs_entropy}). We will show that $\Delta
s_{\rm{tot},\tau}$ satisfies a SSFT for all $\tau$. In
ref.~\cite{Puglisi2006}, the relevance of boundary terms like
$\Delta s_\tau$ has  been pointed out for Markovian processes.

We
investigate the data of the harmonic oscillator described in
sect.\ref{sec:sinusexp}.The probability to compute is the joint
probability $p(\theta(t_i+\tau_n),\dot
\theta(t_i+\tau_n),\varphi)$, where $\varphi$  is the starting
phase $\varphi=t_i \omega_d$. The system is linear, so
$\theta(t_i+\tau_n),\dot \theta(t_i+\tau_n)$  are independent;
thus the joint probability can be factorized into a product.
The   expression of the trajectory dependent entropy is :
\begin{equation}
\Delta s_{\tau_n}= - k_B \ln
\left(\frac{p(\theta(t_i+\tau_n),\varphi)\
p(\dot{\theta}(t_i+\tau_n,\varphi))}{p(\theta(t_i+\tau_n),\varphi)
\ p(\dot{\theta}(t_i+\tau_n,\varphi)}\right)
\label{pendule_heat_trajectory}
\end{equation}

For
computing correctly the trajectory dependent  entropy, we have to
calculate the $p(\theta(t_i),\varphi)$ and
$p(\dot\theta(t_i),\varphi)$  for each initial phase $\varphi$
(see Fig.\ref{fig:trajectories}a).   These distributions turn out
to be independent of $\varphi$ and they  correspond to the
equilibrium fluctuations of $\theta$ and $\dot{\theta}$ around the
mean trajectory defined by $\langle \theta(t) \rangle$ and
$\langle \dot{\theta}(t) \rangle$. The distribution of
$\theta(t_i)$ is plotted in Fig.\ref{fig:trajectories}b), where
the continuous line corresponds to the equilibrium distribution.
Once the $p(\theta(t_i,\varphi)$ and $p(\dot\theta(t_i),\varphi)$
are determined  we compute the "trajectory-dependent" entropy. As
fluctuations of $\theta$ and $\dot{\theta}$ are independent of
$\varphi$  we can average $\Delta s_{\tau_n}$ over $\varphi$ which
improves a lot the statistical accuracy. We stress that it is not
equivalent to calculate first the pdfs over all values of
$\varphi$ --- which would correspond here to the convolution of
the pdf of the fluctuations with the pdf of a periodic signal ---
and then compute the trajectory dependent entropy. The results are
shown in Fig.~\ref{pendule_sinus:pdf}.

In Fig. ~\ref{pendule_sinus:pdf}a), we recall the main results
for the dissipated heat $\Qt=T\Delta s_{\rm{m},\tau_n}$. Its
average value $\langle T.\Delta s_{\rm{m},\tau_n}\rangle$ is
linear in $\tau_n$ and equal to the injected work. The pdfs of
$T.\Delta s_{\rm{m},\tau_n}$ are not Gaussian and extreme events
have an exponential distribution. The pdf of the
"trajectory-dependent" entropy is plotted in
fig.~\ref{pendule_sinus:pdf}b); it is exponential and independent
of $n$. We superpose to it the pdf of the variation of internal
energy divided by $T$ at equilibrium: the two curves match
perfectly within experimental errors, so the
"trajectory-dependent" entropy can  be considered as the entropy
exchanged with the thermostat if the system is at equilibrium. The
average value of $\Delta s_{\tau_n}$ is zero, so the average value
of the total entropy is equal to the average of injected power
divided by $T$. In Fig.~\ref{pendule_sinus:pdf}c), we plot the
pdfs of the normalized total entropy for four typical values of
integration time. We find that the pdfs are Gaussian for any time.

The symmetry functions (eq.\ref{eq:FT}) of the dissipated heat
$S(T\Delta s_{\rm{m},\tau_n}=\Qt)$ and the total entropy $S(T
\Delta s_{\rm{tot},\tau_n})$ are plotted in
Fig.~\ref{pendule_sinus:pdf}d). As we have already seen in
Fig.\ref{fig:S_W_S_Q}, $S(\Qt )$ is a non linear function of
$\Qt=T \ \Delta s_{\rm{m},\tau}$. The linear behavior, with a
slope that tends to $1$ for large time,  is observed only for for
$\Delta s_{\rm{m},\tau_n}<\langle \Delta s_{\rm{m},\tau_n}
\rangle<1$.
For the normalized total entropy, the symmetry functions are
linear with $\Delta s_{\rm{tot},\tau_n}$ for all values of $\Delta
s_{\rm{tot},\tau_n}$ and the slope is equal to $1$ for all values
of $\tau_n$. Note that it is not exactly the case for the first
values of $\tau_n$ because these are the times over which the
statistical errors are the largest and the error in the slope is
large.

For the harmonic oscillator we have obtained that the
"trajectory-dependent" entropy can be considered as the entropy
variation in the system in a time $\tau$ that one would have if
the system was at equilibrium. Therefore the total entropy is the
additional entropy due to the presence of the external forcing :
{\it {this is the part of entropy which is created by  the
non-equilibrium stationary process}}.  The total entropy (or
excess entropy) satisfies the Fluctuation Theorem for all times
and for all kinds of stationary external
torque\cite{Seifert2005,Puglisi2006}. More details on this problem
can be found in ref.\cite{Joubaud2008}.


\section{The non-linear case: stochastic resonance}

\begin{figure}[htbp]
\begin{center}
\includegraphics[width=0.8\linewidth]{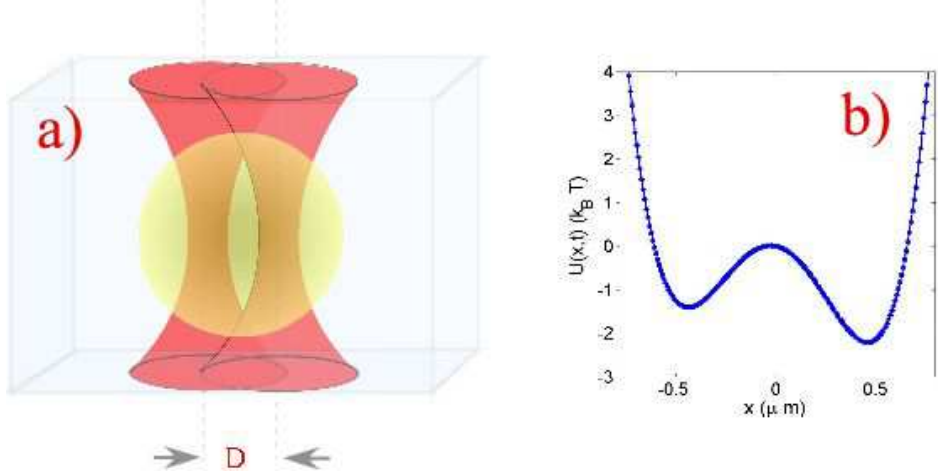}
\caption{a) Drawing of the polystyrene particle trapped by two
laser beams whose axis distance is about the radius of the bead.
b) Potential felt by the bead trapped by the two laser beams. The
barrier height between the two wells is about $2k_BT$. }
\label{fig:two_laser_beams}
\end{center}
\end{figure}

The harmonic oscillator cannot be driven to a non linear regime without forcing
it to such a high level where thermal fluctuations become negligible.
Thus in order to study the non linear effects we change
experiment and we measure the fluctuations of a Brownian particle trapped
in a non-linear potential produced by two laser beams,
as shown in Fig.\ref{fig:two_laser_beams}. It is very well
known that a particle of small radius  $R\simeq 2$~$\mu$m is trapped in the focus of a strongly focused laser beam,
which produces a harmonic potential for the particle, whose
 Brownian motion is confined inside this potential well.
 When two laser beams are focused  at a distance $D \simeq R$,
 as shown in Fig.\ref{fig:two_laser_beams}a) the particle has two
equilibrium positions, i.e.  the  foci of the two beams.  Thermal
fluctuations may
force the particle to move from one to the other.
The particle feels an equilibrium potential
$U_0(x)=ax^4-bx^2-dx$, shown in Fig.\ref{fig:two_laser_beams}b), where
$a$, $b$ and $d$ are determined by the laser intensity and by the
distance of the two focal points. This potential has been computed
from the measured equilibrium distribution of the particle
$P(x)\propto \exp (U_0(x))$. The right left asymmetry of the
potential (Fig.\ref{fig:two_laser_beams}b) is induced by small
unavoidable asymmetries, induced by the optics focusing  the two
laser beams.
 In our experiment the distance
between the two spots is  $1.45~\mu$m, which produces a trap whose
minima are at $x_{min}=\pm 0.45~\mu$m. The total intensity of the
laser is $29~$mW on the  focal plane which corresponds to an
inter-well barrier  energy $\delta U_o=1.8~k_BT$, $ax_{min}^4=
1.8\  k_BT$, $bx_{min}^2=3.6 \ k_BT$ and  $d|x_{min}|=0.44\ k_BT$
(see ref.\cite{Jop_EPL} for more experimental details). The rate
at which the particle jumps from one potentials well to the other
is determined by the Kramer's rate $r_k={1\over \tau_o}
\exp({-\delta U_o \over k_B \ T})$ where $\tau_o$ is a
characteristic time. In our experiment $r_k\simeq 0.3$~Hz at
$300$~K.

\begin{figure}[h!]
\begin{center}
\includegraphics[width=8cm]{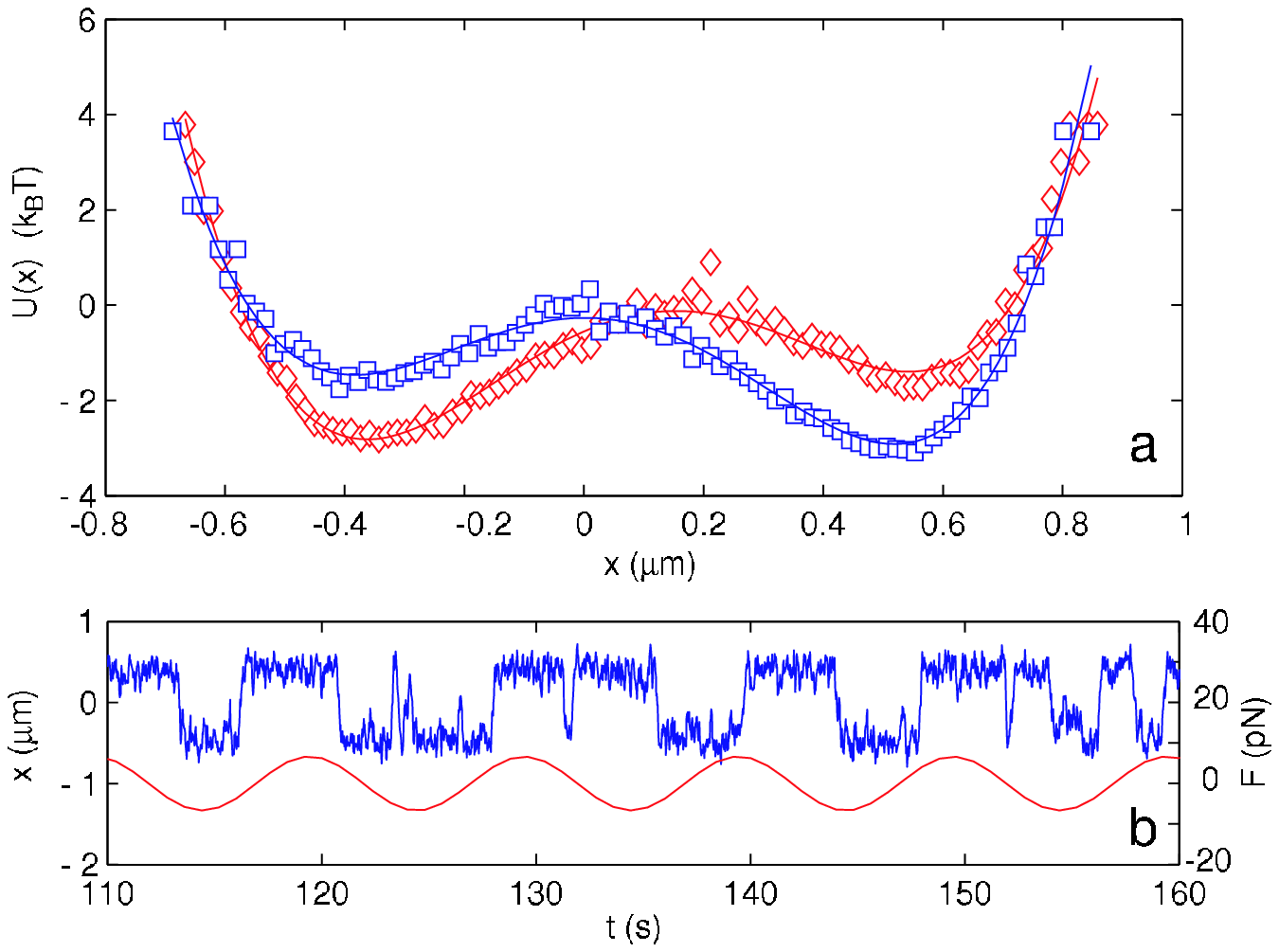}
\includegraphics[width=8cm]{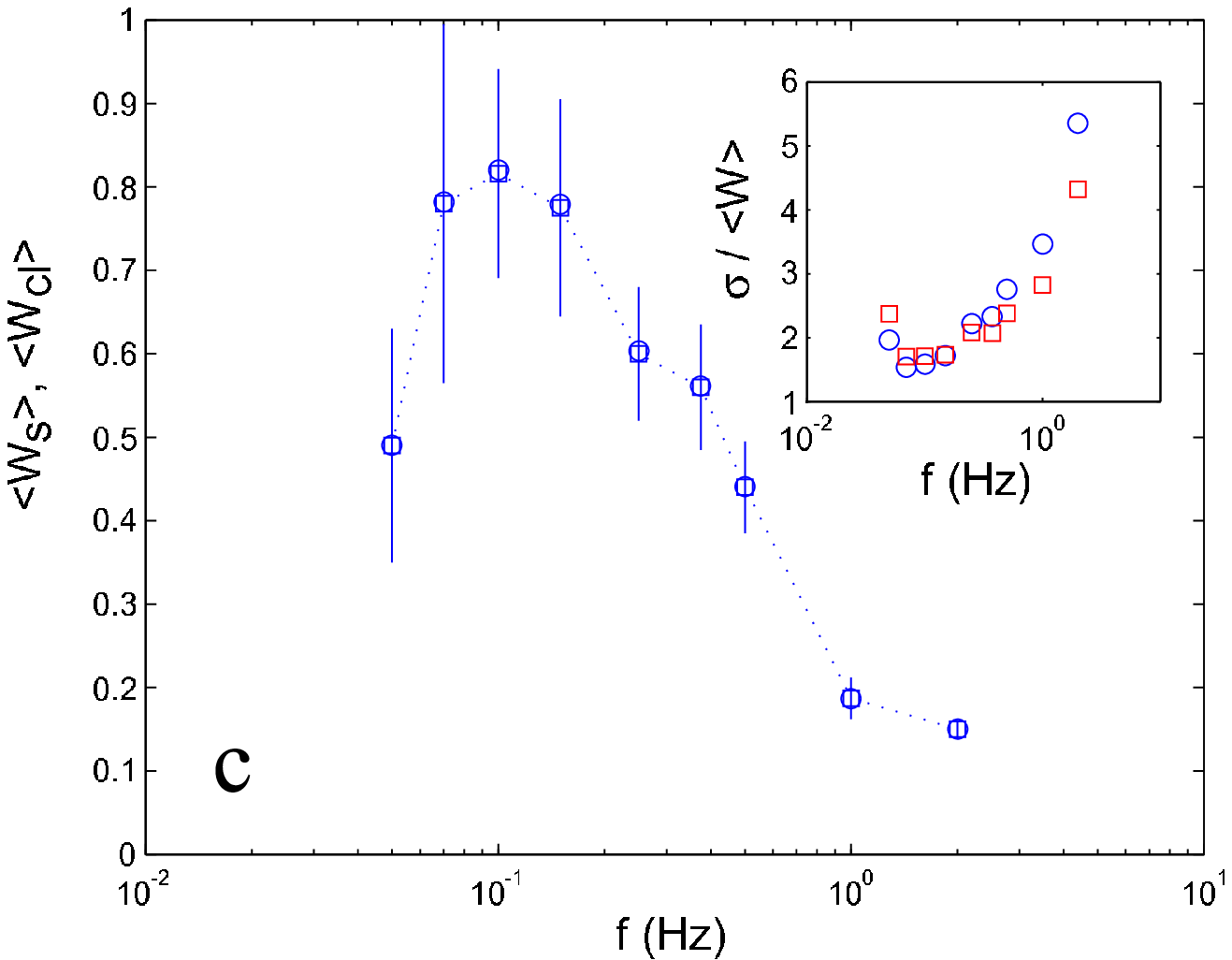}
\caption{a) The perturbed potential at $t=\frac{1}{4 f}$ and
$t=\frac{3}{4 f}$. b) Example of trajectory of the glass bead and
the corresponding perturbation at $f=0.1$ Hz. c) Mean injected
energy in the system over a single period as a function of the
driving frequency.$<W_s>$ $\Box$ and $<W_{cl}>$ $\circ$ coincide
as their mean values are equal within experimental errors. The
error bars are computed from the standard deviation of the mean
over different runs. Inset: Standard deviations of work
distributions over a single period normalized by
the average work as a function of the frequency (same symbols).}
\label{fig:artforcingex}
\end{center}
\end{figure}
To drive the system out of equilibrium we periodically modulate
the intensity of the two beams at low frequency.
Thus the potential felt by the bead is  the following profile:
\begin{equation}
U(x,t)=U_0(x)+U_p(x,t)=U_0+c\ x \ \sin(2 \pi f t),
\end{equation}
with  $c|x_{min}|=0.81~k_BT$. The amplitude of the time dependent
perturbation is synchronously acquired with the bead
trajectory.\footnote{The parameters given here are average
parameters since the coefficients $a$,
$b$ and $c$, obtained from fitted steady distributions at given phases,
vary with the phase ($\delta a/a\approx10\%$, $\delta b/b\approx\delta c/c\approx5\%$).}

An example of the measured potential for $t=\frac{1}{4 f}$ and
$\frac{3}{4 f}$ is shown on the Fig.~\ref{fig:artforcingex}a).
This figure is obtained by measuring the probability distribution
function $P(x,t)$ of $x$ for  fixed values of $c \sin(2 \pi f t)$,
it follows that $U(x,t)=-\ln(P(x,t))$.

The $x$ position of the particle can be  described by a Langevin
equation:
\begin{equation}
   \gamma\dot x=-\frac{\partial U(x,t)}{\partial x}+\eta,
\label{eq:langevin_STR}
\end{equation}
with $\gamma=1.61 \ 10^{-8}$~N~s~m$^{-1}$ the friction coefficient
and $\eta$ the thermal noise delta correlated in in time. When
$c\ne0$ the particle can experience a stochastic resonance
\cite{Benzi}, when the forcing frequency is close to the Kramer's
rate. An example of the sinusoidal force with the corresponding
position are shown on the figure \ref{fig:artforcingex}b). Since
the synchronization is not perfect, sometimes the particle
receives energy from the perturbation, sometimes the bead moves
against the perturbation leading to a negative work on the system.
Two kinds of work  can be defined in this experiments
\cite{Jop_EPL}

\begin{eqnarray}
   W_{s,n}(t)=\int^{t+t_f}_{t}{dt\frac{\partial U(x,t)}{\partial t}}&&  \label{eq:integr_1} \\
   W_{cl,n}(t)=-\int^{t+t_f}_{t}{dt\dot x\frac{\partial U_p(x,t)}{\partial x}}&& \label{eq:integr}\\
\end{eqnarray}
where in this case $t_f={n \over f}$ is a multiple of the forcing
period.
The work  $W_{s,n}$ is the stochastic work (used in Jarzynsky
and Crooks relations \cite{Jarzynski,Crooks,Douarche05}) and $W_{cl,n}$ is
the classical work that will be discussed in this article.
The results on $W_{s,n}$ are quite similar but there are subtle
differences discussed in ref.\cite{Jop_EPL}.

\begin{figure}[h!]
\begin{center}
\includegraphics[width=8.4cm]{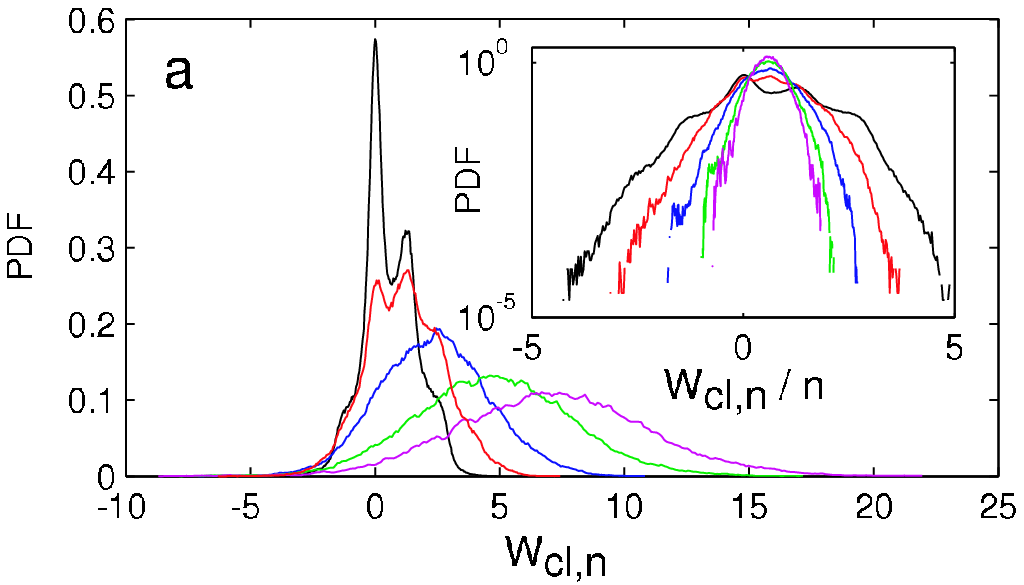}
\includegraphics[width=7cm]{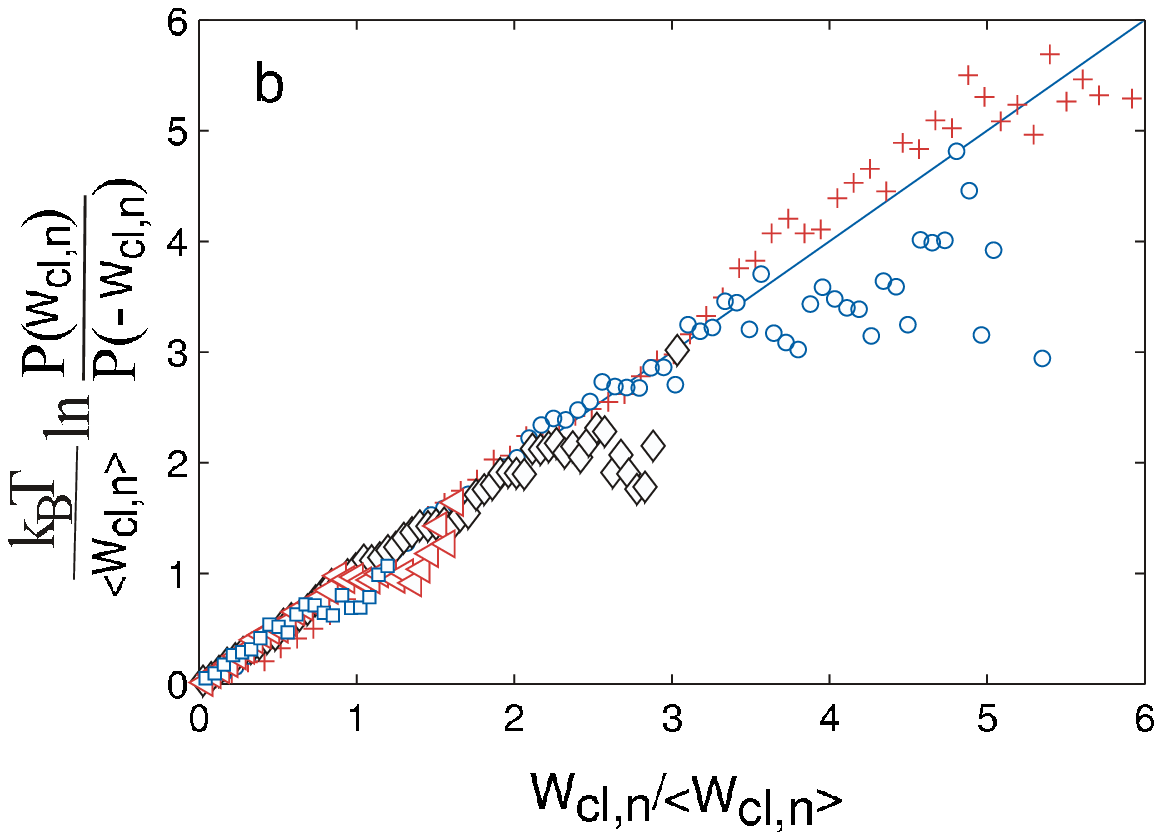}
\caption{a) Distribution of classical work $W_{cl}$ for different
numbers of period $n=1$, $2$, $4$, $8$ and $12$ ($f=0.25$~Hz).
Inset: Same data in lin-log. b)  Normalized symmetry function as
function of the normalized work for $n=1$ ($+$), $2$ ($\circ$),
$4$ ($\diamond$), $8$ ($\triangle$), $12$ ($\Box$).}
\label{fig:wclpdfn}
\end{center}
\end{figure}
We first measure the average work received over one period for
different frequencies ($t_f={1\over f}$ in eq.~\ref{eq:integr}).
Each trajectory is here recorded during 3200~s in different
consecutive runs, which corresponds to 160 up to 6400 forcing
periods, for the range of frequencies explored. In order to
increase the statistics we consider  $10^5$ different $t_o$. The
figure~\ref{fig:artforcingex}c) shows the evolution of the mean work
per period for both definitions of the work. First, the input
average work decreases to zero when the frequency tends to zero.
Indeed, the bead hops randomly several times between the two wells
during the period. Second, in the limit of high frequencies, the
particle has not the time to jump on the other side of the trap
but rather stays in the same well during the period, thus the
input energy is again decreasing when increasing frequency. In the
intermediate regime, the particle can almost synchronize with the
periodical force and follows the evolution of the potential. The
maximum of injected work is found around the frequency $f \approx
0.1$~Hz, which is comparable with half of the Kramers' rate of the
fixed potential $r_K=0.3$~Hz. This maximum of transferred energy
shows that the stochastic resonance for a Brownian particle is a
bona fide resonance, as it was previously shown in experiments
using resident time distributions \cite{gammaitoni95,schmitt06} or
directly in simulations \cite{iwai01,dan05}.
In the inset of Fig.~\ref{fig:artforcingex},
we plot the normalized standard deviation of work distributions
($\sigma/\left<W\right>$) as a function of the forcing frequency.
The curves present a minimum at the same frequency of 0.1~Hz,
in agreement again with the resonance phenomena.

In order
to study FT for stochastic resonance  we choose for the external driving
a frequency $f=0.25$~Hz,  which ensures a good statistic, by allowing the
observation of the system over a sufficient number of periods.
We compute the works and the
dissipation using $1.5 \ 10^6$  different $t$ on time series which
spans about 7500 period of the driving.

We consider the pdf  $P(W_{cl})$ which is plotted in
(Fig.~\ref{fig:wclpdfn}a). Notice that for small $n$ the distributions are double peaked and very complex. They
tend to a gaussian for large
$n$ (inset of Fig. \ref{fig:wclpdfn}a).
On Fig.~\ref{fig:wclpdfn}b), we plot the normalized
symmetry function of $W_{cl,n}$.
We can see that the curves are close to the line of slope one. For
high values of work, the dispersion of the data increases due to
the lack of events.
The slope tends toward 1 as expected by the SSFT.
 It is
remarkable that straight lines are obtained even for $n$ close to
1, where the distribution presents a  very complex and unusual
shape (Fig.~\ref{fig:wclpdfn}a).
We  do  not discuss here the case of $W_{s,n}$  as the behavior is quite similar to that of $W_{cl,n}$ \cite{Jop_EPL}.
The very fast convergency to the asymptotic value of the the SSFT is quite striking in this example.
The measurement are in fully agreement with a realistic model based on the Fokker Planck equations  where the measured values
of $U(x,t)$ has been inserted \cite{Imparato_SR}. This example shows the application of FT in a non-linear case where the distributions
are strongly non-Gaussian.

\section{Applications of Fluctuation Theorems }

The Fluctuations Theorems have several important consequences such
as the Jarzinsky and Crooks
equalities\cite{Jarzynski,Jarzynski1,Crooks}, which are useful to
compute the free energy difference between two equilibrium states
using any kind of
transformation\cite{Douarche05,Douarche04,bustamante,ritort}. The
Hatano-Sasa\cite{Sasa} relations and the recently derived
Fluctaution Dissipation Theorems\cite{Joanny}  are related to FTs
and are useful to compute the response of a NESS using the steady
state fluctuations of the NESS. As we have seen the FT allows the
calculation of tiny amount heat, which can be useful in many
applications in aging systems \cite{Crisanti,Jop_colloid} and
biological systems.

The FTs  for
Langevin systems can   be used to measure an unknown
averaged  power. This idea has been discussed first in the context of
electrical circuits \cite{Garnier} and in ref.\cite{kumiko}
it has been applied for the first time  to the  measure  of the the torque
of a molecular motor. We discuss the method in some details in the next subsection.

\begin{figure}[h!]
\begin{center}
\includegraphics[width=1\linewidth]{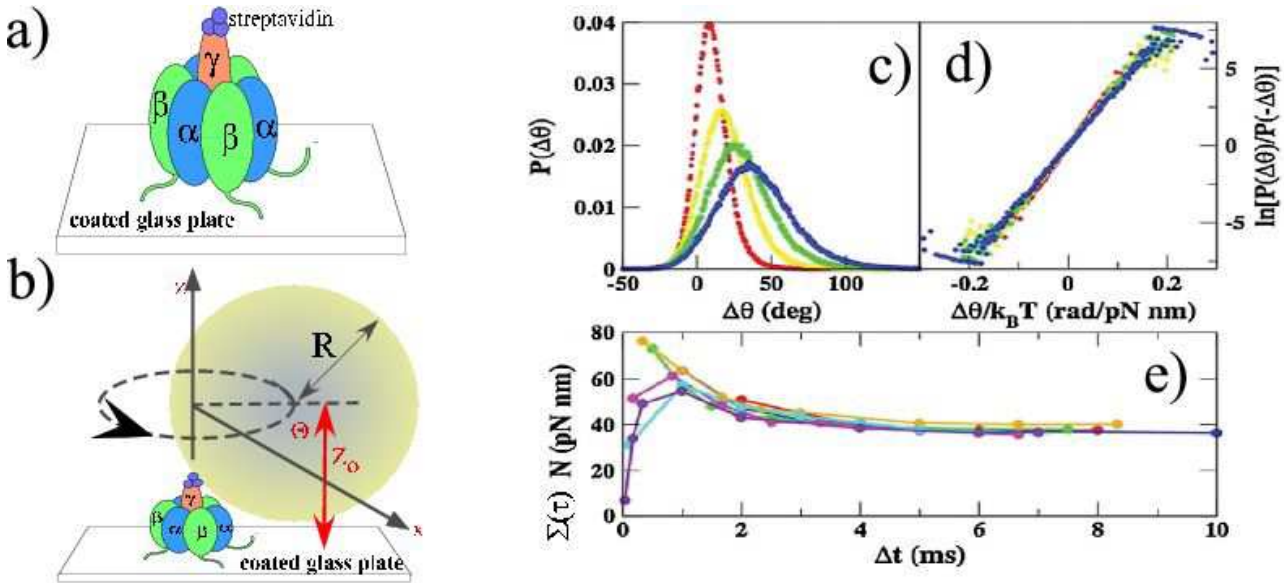}
\caption{
Molecular motor. a) Schematic diagram of a $\rm{F}_1$-ATPase molecular motor
composed by  a rotor $\gamma$ (radius $\sim 1 \ $~nm)  which rotates inside stator  of radius $\sim 5 $~nm formed by three $\alpha \ \beta $ subcomplexes.
Sequential chemical reactions
between the stator and the rotor produce the motion. The $\alpha \beta $ subcomplexes are attached to a suitably coated glass plate. Streptavidin is used to attach to $\gamma$ either actin filaments \cite{Noji} or streptavidin-coated beads \cite{kumiko}.
b) In order to follow the rotation of the rotor with a standard microscope a  streptavidin-coated bead of radius  $\sim 0.5\ \mu$m is glued to $\gamma$ subunit
(drawing not to scale).
The figures c),d) and e) (taken from ref.\cite{kumiko}))
illustrate the results of a measure.
c) Pdfs for several $\tau$  of $\Delta \theta_\tau$.
d) Symmetry function extracted  from the Pdfs of c).
e) Slopes $N \Sigma(\tau) $ of the symmetry function as a function of $\tau$.
The different colors pertain to different experimental conditions.
Notice the convergency to an  unique value of $N$ for large $\tau$. Strictly speaking in this figure the function $\Sigma (\tau)$ keeps into account also the fact that eq.\ref{eq:langevin_motor} is not necessarily valid for short times (see text)} \label{fig:motor}
\end{center}
\end{figure}

\subsection{Measuring the power of a molecular motor}

A molecular rotary motor, as any kind of motor,  is constituted by
a stator and a rotor. The movement of the rotor is provoked by
chemical reactions occurring sequentially between the rotor and
stator. A typical example of bio-motor is the bacterial flagellum.
However it has been shown \cite{Noji} that a single molecule of $\rm{F}_1$-ATPase may
act as a motor composed by a $\gamma$ subunit ( radius $\sim 1 $~nm) which rotates inside a barrel of radius $\sim 5 \ $~nm formed by three $\alpha \ \beta $ subcomplexes (see
fig.\ref{fig:motor}a) for a schematic diagram and ref.\cite{kumiko,Noji} for more details). In experiments,  the $\alpha \beta$
subunit is stuck on a suitably activated glass plate as shown in fig.\ref{fig:motor}a).
The measure of the torque of this motor is important in order to know its efficiency  as a function of the concentration of the chemicals
contained in the liquid surrounding it. The typical size of
this  molecular motor is several nanometers and the moving unit
is too small to be observed with an optical microscope. Therefore
to measure the torque of $\rm{F}_1$-ATPase  motor a streptavidin-coated bead of radius $0.5 \ \mu$m is glued on the
subunit $\gamma$, and the motion of this bead is followed by a
standard microscope as sketched in Fig.~\ref{fig:motor}b). The
motion of the bead occurs on a torus and $\theta$ is the
coordinate of the motion along the torus.  The time evolution of
$\theta$ can be described by a Langevin equation:

\begin{equation}
   \Gamma \ \dot \theta=N+\xi,
\label{eq:langevin_motor}
\end{equation}

where $N$ is the torque of the motor, $\Gamma$ the viscous
dissipation of the fluid surrounding the bead and  $\xi$ the
thermal noise. In order to estimate $N$, it is in principle enough
to measure the mean angular velocity $<\dot \theta >$ of the bead
and obviously from eq.\ref{eq:langevin_motor} one gets:
\begin{equation}
   N=\Gamma \ <\dot \theta>
\label{eq:torque_motor}
\end{equation}
However the procedure  is not so simple because in order to
estimate $N$ one has to know the exact value of $\Gamma$, which is a
function of the viscosity of the fluid, the radius $L$ of the bead
path, the radius $R$ of the bead, the shape of the bead  and the
distance $Z_o$ of the bead from the surface of the glass  plate  where the
experiment is performed (see Fig.~\ref{fig:motor}b). The variable
$Z_o$ is certainly the most
difficult to be determined experimentally. Therefore using
eq.\ref{eq:torque_motor} the error on $N$ can be really very
large. There is instead another  method proposed in ref.\cite{Garnier} for electric circuits and first used for measurements in bio-motors in ref.\cite{kumiko}. This method is based on work fluctuations
and  is much more precise. To apply it,  we suppose that $N$ is
constant. This is a quite reasonable hypothesis for certain regimes of $\rm{F}_1$-ATPase motors.  We compute the
work $W_\tau$ performed by the motor in a time $\tau$:
\begin{equation}
   W_\tau(t)=\int_t^{t+\tau} \ N \ \dot \theta (t) \dd t= N\
   \Delta \theta_\tau
   \label{eq:work_motor}
\end{equation}
where $\Delta \theta_\tau= \theta(t+\tau)-\theta(t)$ and we have
used the fact that $N$ is constant.  In previous section we have
seen that Langevin systems satisfy the SSFT, which we now  apply to
$W_\tau$. Using eq.\ref{eq:work_motor} and the constancy of $N$
SSFT for the molecular motor reads:

\begin{eqnarray}
\ln \left( {P(\Delta \theta_\tau) \over P(-\Delta \theta_\tau)}
\right) = \Sigma (\tau) \ N \  {  \Delta \theta_\tau  \over k_BT}
\ \  \ \  \ \  \ \  \ \  \ \  \ \  \ \  \rm{with} \ \  \Sigma(\tau) \rightarrow 1 \ \ \rm{for} \ \ \tau
\rightarrow  \infty  \label{eq:SSFT_motor}
\end{eqnarray}

This equation is quite interesting because the value of $N$ can be
determined only by the measure of the fluctuations of $\Delta
\theta_\tau$. Indeed plotting $\ln ( {P(\Delta \theta_\tau) /
P(-\Delta \theta_\tau)})$ as a function of $\Delta \theta_\tau /(
k_BT)$ we notice that the slope of the straight lines is $\Sigma
(\tau) \ N$. Therefore studying the asymptotic value of this slope
for large $\tau$ one can determine $N$. It is interesting to note
that in this case the knowledge of $\Gamma$ is not needed. This
technique has been recently applied to molecular motor in
ref.\cite{kumiko} and their main results are plotted in
Figs.~\ref{fig:motor} c),d),e). The relevant parameter $\Sigma
(\tau) \ N$, extracted from the pdf of $\Delta \theta_\tau$
(Fig.~\ref{fig:motor} c) and the  symmetry function
(Fig.~\ref{fig:motor} d), is plotted in Fig.~\ref{fig:motor}e) \
\footnote{It has to be stressed that in this specific case $\Sigma
(\tau)$ keeps into account the fact that for short time
eq.\ref{eq:langevin_motor} is not a good approximation for the
dynamics of the motor and
eqs.\ref{eq:work_motor},\ref{eq:SSFT_motor} apply only for long
time}. We see a clear convergence to a unique value and one  gets
a very precise estimation of the torque of the molecular motor
independently of the size and shape of the bead glued to $\gamma$
unit of the $\rm{F}_1$-ATPase motor. This is a very specific and
interesting example of the possible applications of FT.

\section{The chaotic systems\label{section:dynamical_system} }

In previous sections we have studied the probability of the
instantaneous negative entropy production rates within  the
context of the FTs for stochastic systems, where the  fluctuations
are produced by the coupling with a thermal bath. In
sec.~\ref{sec:gaussian_FT} we have seen that when the  energy
injected into the system is larger than $100 \ k_BT$ the
probability of these negative events is very small and  the time
needed to observe them becomes extremely long.  In other words the
role of thermal fluctuations becomes negligible.

However in the introduction we have shown that  instantaneous
negative
entropy production rates {  can be observed in chaotic systems
such as, for example, turbulence and  granular media, where the fluctuations are produced
 by the non linear
interactions of many degrees of freedom.  We have also pointed out
that for chaotic systems the amount of injected energy is order of
magnitudes  larger than $k_BT$ and of course thermal fluctuations
do not play any role in the fluctuating dynamics. The question
that we want to analyze in this section is whether we can apply in
these  systems the FTs
defined in sec.\ref{section_FT} for stochastic systems, eqs.\ref{eq:FT}-\ref{eq:FTs}.
For a dissipative chaotic  system one could imagine to replace
$k_BT$, in eqs.\ref{eq:FT}-\ref{eq:FTs}, with a characteristic
energy $E_c$ which  keeps into account the relevant energy scales
of the  system fluctuations. However the definition of this
relevant energy scale can be in general difficult and even
impossible, because it may depend on the observable and on the
kind of forcing.  Thus the approach of introducing an $E_c$ is not
very useful to compare the experimental results with the proof
given for dynamical systems \cite{GallavottiCohen95}}. Indeed in
this case the theorem considers a quantity :
\begin{eqnarray}
y_\tau= \frac {\sigma_\tau} {<\sigma_\tau>} = \frac {\int_t^{t+\tau} \sigma(t) \ \dd t}
{ <\sigma_\tau> }
\label{eq:y_tau}
\end{eqnarray}
where $\sigma(t)$ is the instantaneous phase
space contraction rate, $\sigma_\tau$ the integral of $\sigma$ on
a time $\tau$ and $<\sigma_\tau>$ the mean of $\sigma_\tau$. Three
hypothesis has been done on  the dynamical system which must be  :
a) dissipative, b) time reversible c) Anosov \footnote{ For a
precise definition see refs.\cite{Anosov,Zamponi_comment}. Roughly
speaking this property ensures that the system is chaotic and
that on the attractor there are no regions of finite volume that
do not contain points.}
\begin{eqnarray} {1 \over
\tau} \ln \frac{P(y_\tau)}{P(-y_\tau)} =   \ {<\sigma> }  \
{{y_\tau}}  + O(1/\tau) \quad   \rm{for} \quad \tau \rightarrow
\infty \label{FT_dynamical}
\end{eqnarray}
where $<\sigma>$ is the mean phase space contraction rate, which
has the dimension of $1/t$. In this equation the relevant variable
is the phase space contraction rate which has been identified as
the entropy production rate\cite{GallavottiCohen95}. The phase
space contraction rate is a global variable of the system but an
extension of the theorem for local variables has been done in
refs.\cite{Gallavotti_Local,Ayton}. One reason for developing
local FT is that global fluctuations are usually not observable in
macroscopic systems, as a consequence it is important to
understand whether a local measurement is representative of the
dynamics.
Eq.\ref{FT_dynamical} has
been tested in several numerical simulations (see for example
ref.\cite{Rondoni2007,Zamponi_comment} for a review), here we want to focus on
experiments.

\subsection{Experimental test}
The test of eq.\ref{FT_dynamical} in experiments is extremely
useful to analyze several important questions. The first one is
whether eq.\ref{FT_dynamical} may have a more general validity
independently of the restrictive hypothesis done to prove it.
Indeed  the hypothesis b) is never satisfied in  real systems and
the hypothesis c) does not necessarily  apply to all of them. Thus
in general we do not even know whether eq.\ref{FT_dynamical} can
be applied in the experimental system under study. The second
question concerns the choice of the observable. Indeed the direct
measure of the phase space contraction rate is not possible and
one has to rely upon  the measure of another observable  usually
the  energy $\Wt$  injected into the system by the external forces
in a time $\tau$. In other words one is making the important
hypothesis  that $y_\tau$, defined in eq.\ref{eq:y_tau} in terms
of $\sigma(t)$ is equivalent to $x_\tau=\Wt/<\Wt>$. This
hypothesis, that is not necessarily valid,  is the second question
that one would like to address in experiments.
The third question is related to the estimation of the  prefactor
$<\sigma>  $ in the right hand side of eq.\ref{FT_dynamical}.
This prefactor, which is a function of the Lyapunov exponents,
is very difficult to estimate in an experimental system.
Finally the last question concerns the relevance
of a local observable to characterize the dynamics of the system.

There are not many experiments where these questions   have been
analyzed in some details. In several experiments
\cite{Ciliberto_turbulence,Ciliberto98} only the linearity in
$x_\tau$ of the symmetry function $\rho(x_\tau)=(1/\tau) \ln
\left( {P(x_\tau)}/{P(-x_\tau)} \right) $  has been checked,
which, for the reasons discussed in the previous paragraph, is
only a partial test.  {  For example three experiments
have tried to give an answer to the question
of the prefactor}. Two of these experiments are
performed in granular media \cite{Menon,Joubaud_granular}
and the third on mechanical wave turbulence in a metallic plate\cite{Cadot}.
We will not describe in details the experiments here but we will
comment the main results.

\subsubsection{\it Granular media}
The two experiments of refs.\cite{Menon,Joubaud_granular}
consist of  diluted granular media strongly shaken by a vibrator, but the measured quantities  are not the same.
{  In ref.\cite{Menon} the authors measure the fluctuations of the energy flux in a subvolume of the system. Instead ref.\cite{Joubaud_granular} the work done by an external force on a ratchet inside the granular media is measured.}
They both find that although
the system is not thermal the stochastic version of SSFT (eq.\ref{eq:FT}) holds,
provided that $k_BT$ is replaced by a characteristic energy $E_c$:

\begin{eqnarray}
\ln \frac{P(x_\tau)}{P(-x_\tau)}
=   \ \frac {<X_\tau>}  {E_c}  {{x_\tau}}  + O(1/\tau).
\label{FT_granular}
\end{eqnarray}
with $x_\tau=X_\tau/<X_\tau>$, {  $X_\tau$ is the integral of energy flux in ref.\cite{Menon} and the work $\Wt$ performed by an external force on a ratchet in ref.\cite{Joubaud_granular}.
It must be pointed out that in both experiments of refs.\cite{Menon,Joubaud_granular} the energy $E_c$ has been
measured independently. In ref.\cite{Menon} is found that $E_c$ is about 5 times larger than
the kinetic energy $K_E$ of the shaken granular medium for all the values of the control parameters used in the experiment.
The fact that  $K_E$ and $E_c$ have the same dependence on the control parameter have  been interpreted considering that the vibrator
injects into the system the amount of energy  lost in the
collisions but once excited in a NESS the granular medium behaves like a thermal bath for
 the measured observable.  However this interpretation is not necessarily correct. Indeed
for the experiment of ref.\cite{Menon} it as been shown in a
numerical simulation \cite{Puglisi} that for the quantity measured
in this experiment FT does not apply for the large deviations
$x_\tau$ because the symmetry function $S(x_\tau)$ becomes
non-linear for large $x_\tau$. This discrepancy between theory and
experiment is obviously coming from the fact that experimentally
the very large deviation are difficult to be measured, thus the
non-linear part of $S(x_\tau)$ cannot be observed. However the
experiment of ref.\cite{Menon} is certainly interesting because is
the first where the question of the prefactor has been analyzed
experimentally. In contrast for the experiment of
ref.\cite{Joubaud_granular} is observed that $E_c= K_E (1+\alpha)/2$ where $\alpha$ is the
 restitution coefficient of the grains. It is interesting to notice that
  $K_E (1+\alpha)/2$ is the temperature of an intruder inside a
   diluted granular gas as it has been found in  theoretical models
\cite{Sarracino,Costantini,vdb07}.
These two examples of comparison between numerical and
experimental results show the difficulty of interpreting the
experimental results on FT and the importance of verifying them in
a precise theoretical framework. Where does the difference between
the experiments of  ref.\cite{Menon} and
ref.\cite{Joubaud_granular} come from ? The answer can be found on
the fact that in the two experiments  two different quantities are
measured. Indeed in sect.\ref{section_FT} we have seen that, even
for stochastic systems, the fluctuations of $\Wt$, $\Qt$  and
$\Delta s_{\rm{tot},\tau_n}$ behave differently within the context
of FT. This can be  more complex for granular media  and it will
be useful to give more insight on this point. Furthermore even in
cases where a description in terms of $Ec$ applies, comparing
eq.\ref{FT_granular} with eq.\ref{FT_dynamical}
 a question that  arises naturally  is whether
$<\Wt> / E_c$  is a good estimation of $<\sigma> \ \tau$.
This is an important question which will be interesting to analyze in the future.}

\subsubsection{\it Mechanical waves}

\begin{figure}[h!]
\begin{center}
\includegraphics[width=0.9\linewidth]{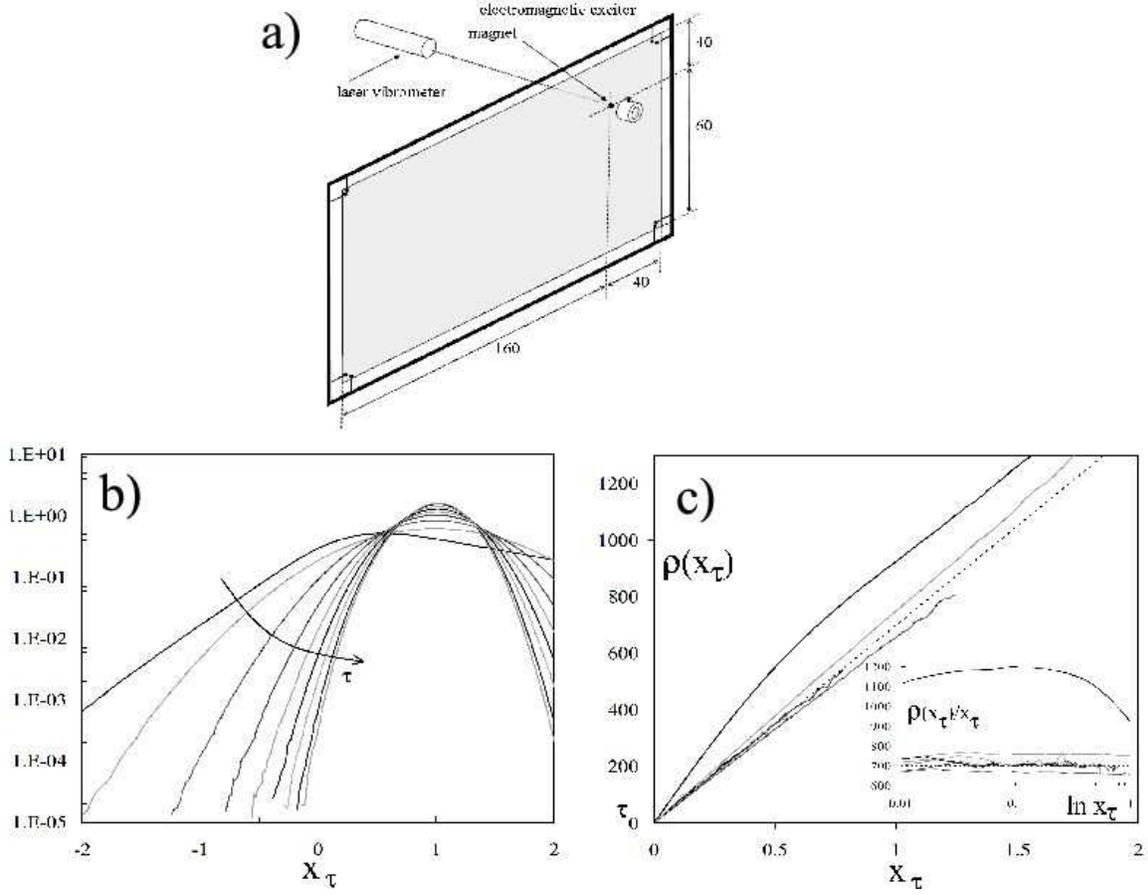}
\caption{ Mechanical waves in a metallic plate (from
ref.\cite{Cadot}).   a) Schematic diagram of the experiment. A
steel plate is suspended to the frame. Dimensions are in cm. The
electromagnetic exciter produce et local forcing of the plate
vibrations. A laser vibrometer measures the normal velocity at the
excitation point. b)and c) Results on the periodic forcing. (b)
Pdfs of the injected power on the time durations of $\tau$ for
$\tau= 3.5 ms, \ 6.5 , \ 13.5 , \ 20 , \ 26.5 , \  33.5,\ 40 ,\
47.5 , \ 52 ms$ )  (c)  Functions $\rho(x_\tau)=(1/\tau) \ln
\left( {P(x_\tau)}/{P(-x_\tau)} \right) $  obtained from the Pdfs
of (b). Inset: compensated value $\rho(x_\tau)/x_\tau$ in a semi
log plot. In (c) the dashed line corresponds to a linear law of
slope $<\sigma> = 700 Hz$.} \label{fig:Cadot}
\end{center}
\end{figure}

In the experiment on mechanical waves\cite{Cadot}, a metallic  plate
is set into a chaotic state of wave turbulence by a periodic
local forcing at $75$~Hz. (see Fig.~\ref{fig:Cadot}a). The chaotic
dynamics is produced by the non-linear interaction of the
oscillatory modes of the plate. The authors measure the local
force and displacement (see Fig.~\ref{fig:Cadot}a) and compute the
work $\Wt$  done on a time $\tau$ by the external force which
excites the vibrations of the plate. They find that the pdfs of
$x_\tau=\Wt/<\Wt>$ are strongly non-gaussian (see
Fig.~\ref{fig:Cadot}b). From these pdfs they compute the symmetry
function $\rho(x_\tau)$ which is plotted  (see
Fig.~\ref{fig:Cadot}c)   as a function of $x_\tau$. We see that in
spite of the fact that the pdf are not Gaussian the function
$\rho(x_\tau)$ (Fig.~\ref{fig:Cadot}c)  converges to a unique
straight line for large $\tau$ as predicted by FT. From
eq.\ref{FT_dynamical} the slope of this straight line is
$<\sigma>$, which the authors  can estimate independently by
measuring the relaxation time of the vibrational modes. They find
that the values estimated with the two methods (FT and the
relaxation time)  are very close and within experimental errors.
This result is quite interesting and it is probably the only
experiment  where a direct test of eq.\ref{FT_dynamical} has been
done. Certainly the errors of this comparison are very large but
this kind of tests are useful to understand  in some details the
applications of FTs  to chaotic systems.

\section{Summary and concluding remarks}

In this paper we have reviewed several experimental results on the fluctuations of injected and dissipated power in out of equilibrium systems. We considered the two cases when the fluctuations are produced by the coupling with the heat bath (stochastic systems) and when they are
produced by the non linear interactions of many degrees of freedom
{  (chaotic systems)}.
 We have seen that in both cases we  observe that the external
 forces may produce a negative work because of fluctuations.
 The probability of these negative events has been analyzed in the framework
 of fluctuation theorem.

We have mainly discussed the stochastic systems described by Langevin equations,
both with harmonic and unharmonic potential. We have seen that  injected and dissipated
power present different behaviors. FTs are valid for any value of $\Wt$ whereas
can be applied only for $\Qt\ < <\Qt> $ in the case of the heat.
We have also seen the the finite time corrections to SSFT depend on the driving
and on the properties of the system. We have introduced the total entropy,
which takes into account only the entropy produced by the
external forces neglecting the the equilibrium fluctuations.
For the total entropy  FTs are valid for all the times.
We discussed the applications of FTs to extract important physical properties
of a stochastic system.
Thus one may conclude that for Markovian
systems driven by a deterministic force the applications
 of FT does not present any major problems and can be safely applied.
 The case of random driving has been recently discussed
 and several problems may arise when  the variance
 of the driving become larger than the fluctuations induced
 by the thermal bath.
 We have not discussed this problem but
 an analysis of this specific case can be found in refs.\cite{Solano_AFM,Baule,Bonaldi}.

{  Finally we discussed the applications of FT to
chaotic systems. The experimental test is in this case very
important and useful because many questions can be asked on the
system under study which does not necessarily verify all the
theoretical hypothesis. One has to say that
in the case of non-Gaussian statistics even the linearity
of the symmetry function can be an interesting result.
However we pointed out that, for a real comparison with theory,
the  difficulty is to estimate of the
prefactor of eq.\ref{FT_dynamical}
by  an independent measurement.
Only a few experiments have addressed this point in some details,
 but many problems remain open and it seems to be difficult to find
 a general behavior for chaotic systems as
 for the case of stochastic ones.}

{\bf Acknowledgements}

Several results reviewed in this article has been obtained in
collaboration, with L. Bellon, R. Gomez-Solano, N. Garnier, P.
Jop. We acknowledge  useful discussions  and collaboration during
the past ten years with D. Andrieux, E. Cohen, R. Chetrite, G.
Gallavotti, P. Gaspard, K. Gawedzky and  A. Imparato. We thank A.
Puglisi and E Trizac for useful comments on FT in granular media.
We are grateful to  H. Noji, K. Hayashi and A Boudauoud for having
allowed us to reproduce the figures of their articles in the
present one.

\end{document}